\documentclass[11pt,a4paper]{article}

\RequirePackage[l2tabu, orthodox]{nag}
\usepackage[T1,T2A]{fontenc}
\usepackage [utf8]{inputenc}

\usepackage{jheppub}
\usepackage{amsthm,amssymb,amsmath,epic,float}
\usepackage{rotating,epsfig,indentfirst,array,varioref}
\usepackage{appendix,tikz,mathtools}
\usepackage{setspace}
\usepackage{tikz}
\usetikzlibrary{decorations.pathreplacing}
\usetikzlibrary{calc}
\usepackage{enumitem}
\usepackage{hyphenat}
\usetikzlibrary{positioning}
\usepackage{graphicx}  
\usepackage{adjustbox}

\usepackage{arydshln}
\usepackage{cancel}

\usepackage{subfig}
\usetikzlibrary{intersections}
\pgfdeclarelayer{background}
\pgfsetlayers{background,main}

\usepackage{tikz,pgf}
\usetikzlibrary{shapes}
\usetikzlibrary{calc}
\usetikzlibrary{decorations.pathmorphing}
\usetikzlibrary{decorations.pathreplacing,shapes.misc}
\usetikzlibrary{positioning}
\usetikzlibrary{arrows}
\usetikzlibrary{decorations.markings}
\usetikzlibrary{shadings}
\usetikzlibrary{intersections}

\def\be{\begin{equation}}
\def\ee{\end{equation}}

\def\be{\begin{equation}}
\def\en{\end{equation}}
\def\ber{\begin{eqnarray}}
\def\enr{\end{eqnarray}}

\newcommand{\pd}{\partial}

\newcommand{\br}[1]{{\overline{#1}}}

\def\<{\left(}
\def\>{\right)}

\usepackage{jheppub}
\makeatletter
\def\@fpheader{\vspace{-.1cm}}
\makeatother

\title{\boldmath $p \to \infty$ limit of tachyon correlators in $(2,2p+1)$ minimal Liouville gravity from classical Liouville theory}

\author[a,b]{A.~Artemev}

\affiliation[a]{Landau Institute for Theoretical Physics, 142432 Chernogolovka, Russia}
\affiliation[b]{Skolkovo Institute of Science and Technology, 121205, Moscow, Russia}

\emailAdd{artemev.aa@phystech.edu}

\abstract{
Previously it was suggested, motivated by correspondence with JT gravity, that tachyon correlators in $(2,2p+1)$ minimal Liouville gravity (MLG) in the $p\to \infty$ (semiclassical) limit should be interpreted as moduli space volumes for constant curvature surfaces with conical defects. In this work we propose that these volumes are associated with Kähler metrics on moduli spaces introduced by Zograf and Takhtajan, for which the classical Liouville action is a Kähler potential. We check this proposal by numerical calculation of these Kähler metrics and associated volumes for the simplest example of genus 0 surface with 4 conical defects, using conformal field theory. A peculiar property of MLG correlators is proportionality to number of conformal blocks in a certain region of parameter space; in a particular limiting case, we check this property for the volumes following from classical Liouville action and thus provide an analytic confirmation of our proposal. }


\begin{document} 
\maketitle
\flushbottom

\section{Introduction}
The topic of 2-dimensional quantum gravity is intensively studied since the 1980s (\cite{Knizhnik:1988ak}, \cite{DISTLER1989509} and numerous other works) and is quite developed. Different approaches and models in this area are deeply and intricately connected. One of such connections is the one between minimal Liouville gravity (MLG) and Jackiw-Tetelboim gravity (JT, see \cite{mertens2022solvable} for comprehensive review); it is believed that JT gravity can be thought of as semiclassical limit of $(2,2p+1)$ MLG. It was first noted in \cite{saad2019jt}, where matrix model description of JT gravity was developed, and consequently verified by explicit calculations of different observables in both field-theoretic (e.g. \cite{mertens2021}) and matrix model approaches. Of particular interest are observables involving integration over moduli of the surface; a well-studied example in MLG are the so-called tachyon correlation numbers (see e.g. \cite{Zamolodchikov:2005fy}, \cite{Belavin:2005jy}). In \cite{turiaci2021}, semiclassical (or JT) limit of tachyon correlators was considered from the matrix model side. For these an interpretation as ``volumes of moduli spaces for constant curvature surfaces with conical defects'' was proposed. A few other arguments in support of this were given in \cite{Artemev_2022}.

A purpose of this article is to propose a particularly natural  in this context connection of these results to certain metrics and volumes on the moduli space, already known in the literature. They were introduced by Zograf and Takhtajan in the works \cite{Zograf_1988}, \cite{takhtajan2001hyperbolic} on classical Liouville theory. The most ``hands-on'' definition of these is that they are Kähler metrics, potential for which is the classical Liouville action. For brevity we will later refer to these as ZT metrics, to distinguish from both what is known as Takhtajan-Zograf metrics in mathematical literature and Weil-Petersson metrics, reserving this name for metrics on moduli space of constant negative curvature (hyperbolic) surfaces with punctures(cusps) and/or geodesic boundaries.

The structure of this paper is as follows. In section \ref{sec2}, we introduce notations and necessary facts from classical and quantum Liouville field theory and minimal gravity and define the object of our study. Section \ref{sec3} is devoted to description and results of numerical method for calculating moduli space volumes associated with ZT metrics for a one-parametric family of geometries with 4 conical defects on a sphere; we also discuss the limitations and applicability of this method. In section \ref{sec4}, on a few examples we analytically confirm that ZT volumes conform to a certain known property of semiclassical MLG tachyon correlators. We conclude in \ref{sec5} with some discussion of related questions that would be interesting to study further.
\section{Preliminaries} \label{sec2}
\subsection{Quantum and classical Liouville CFT} 
Liouville conformal filed theory is a 2-dimensional CFT which has the action of the form
\begin{equation}
    A_L = \int d^2x\,\sqrt{\hat{g}}\left(\frac{1}{4\pi}\hat{g}^{ab} \pd_a \phi \pd_b \phi + \mu e^{2b\phi} + \frac{Q}{4\pi} \hat{R} \phi \right)
\end{equation}
$\hat{g}$ is a reference metric that we introduce to give theory a covariant form. We will only consider Liouville theory on the sphere in this paper. Parameter $Q = b + b^{-1}$; the Virasoro central charge of the theory is then given by $c_L = 1 + 6 Q^2$. $\mu$ is the  parameter of the theory called the ``cosmological constant'' in this context; dependence of the correlators on $\mu$ is fixed, e.g. by noticing that $\mu$ can be put to one by shifting $\phi \to \phi - \frac{1}{2b} \log \mu$.

Holomorphic stress-energy tensor of the theory is $T = - (\pd \phi)^2 + Q \,\pd^2 \phi$. Exponential operators $V_a = e^{2a \phi}$ are primary fields of the model of dimension $\Delta^L_a = a (Q-a)$. Operators $V_a$ and $V_{Q-a}$ are identified up to a factor $R_L(a)$ called the ``reflection coefficient''.

As usually the case in CFT, in Liouville theory there are the so-called ``degenerate'' fields. They correspond to fields $V_{m,n} \equiv V_{a_{m,n}}$ with
\begin{equation}
a_{m,n} = - b^{-1} \frac{(m-1)}{2} - b \frac{(n-1)}{2} \label{lioprim}
\end{equation}
These are the primary fields (Virasoro highest vectors) with a descendant on level $mn$ which is annihilated by positive part of the Virasoro algebra and is a highest vector by itself (in other words, the corresponding Vermat module has a submodule). To work only with irreducible representations, in CFT such submodules are usually decoupled by putting its highest vector to zero. Decoupling conditions can be written as $D_{m,n} V_{m,n} = 0$, where $D_{m,n}$ is a polynomial of Virasoro generators $L_{-k}$ of degree $mn$; in particular, $D_{1,2} = L_{-1}^2 + b^2 L_{-2}$. From this follows the celebrated BPZ equation \cite{bpz} for a 4-point correlator with degenerate field:
\begin{equation}
\left[\frac{\pd^2}{\pd z^2} + b^2 \left(\sum \limits_{k=1}^3  \left(\frac{\Delta_k}{(z-z_k)^2} - \frac{1}{z-z_k} \frac{d}{dz} \right) - \sum \limits_{i<j} \frac{\Delta_{1,2}+\Delta_{ij}}{(z-z_i)(z-z_j)} \right) \right] \langle V_{1,2}(z) \prod \limits_{i=1}^3 V_{a_i}(z_i)  \rangle = 0 \label{bpz}
\end{equation}
 
An expression for the three-point functions (or structure constants) of Liouville theory $C_L(a_1, a_2, a_3)$ was first proposed by two groups of authors \cite{dornotto1994} and \cite{zamzam1996}. It is called Dorn-Otto-Zamolodchikov-Zamolodchikov (DOZZ) three-point function. The formula is
\begin{equation}
C_L(a_1, a_2, a_3) = (\pi \mu \gamma(b^2) b^{2-2b^2})^{(Q-a)/b} \frac{\Upsilon_b(b) \Upsilon_b(2a_1)\Upsilon_b(2a_2)\Upsilon_b(2a_3)}{\Upsilon_b(a-Q) \Upsilon_b(a-2a_1) \Upsilon_b(a-2a_2)\Upsilon_b(a-2a_3)} \label{liouv3pf}
\end{equation}
where $\Upsilon_b(x)$ is a certain special function that we will not be discussing in detail. We cite two important properties of it: the shift relations 
\begin{align*}
    & \Upsilon_b (x+b) = \gamma(bx) b^{1-2bx} \Upsilon_b(x)\\
    & \Upsilon_b (x + b^{-1}) = \gamma(x/b) b^{2b/x-1} \Upsilon_b(x)
\end{align*}
with $\gamma(z) = \frac{\Gamma(z)}{\Gamma(1-z)}$ and the fact that this function has zeroes for $x= - m b - \frac{n}{b}$ and $x = Q + \frac{m}{b} + n b$ for $m,n$ non-negative. 

The OPE $V_{a_1}(x) V_{a_2}(0)$ in Liouville CFT is most simply written when $a_1, a_2$ lie in the so-called ``basic domain'' defined by
\begin{equation}
    \left|\frac{Q}{2} - \text{Re }a_1\right| + \left|\frac{Q}{2} - \text{Re }a_2\right| < \frac{Q}{2} \label{basicdom}
\end{equation}
In this case we write
\begin{equation}
    V_{a_1}(x) V_{a_2}(0) = \int \limits_{-\infty}^{\infty} \frac{dP}{4\pi} C_L(a_1, a_2, \frac{Q}{2}- i P) (x \br{x})^{\Delta^L_{Q/2 + i P} - \Delta^L_{a_1} - \Delta^L_{a_2}} [V_{Q/2+iP} (0)]
\end{equation}
The fields that appear here are parametrized by one real number $P$ instead of complex $a$ --- only operators with $a = \frac{Q}{2} + i P,\,P \in \mathbb{R}$, correspond to normalizable, or ``physical'', states of the theory \cite{seibergnotes}. This is why we sum over these states only in the OPE. 
When parameters of the correlators are not in this domain, some poles of the structure constants may cross the contour over which we integrate. In order to keep the analyticity in parameters, in such case we should either deform the contour of integration over $P$ or (equivalently) keep the contour the same but explicitly add contributions from these poles which are referred to as ``discrete terms''.

Liouville theory has a nice semiclassical limit which is obtained when $b \to 0$ (or $c \to \infty$). After rescaling the field $b\phi(x) = \varphi(x)$, in terms of $\varphi$ the action has a large prefactor $\sim 1/b^2$. Functional integral is then saturated by a saddle point --- solution of classical Liouville equation
\begin{equation}
\pd \br{\pd} \varphi_{cl} = \Lambda e^{2\varphi_{cl}},\,\Lambda = 4 \pi \mu b^2
\end{equation}
Function $\varphi_{cl}$ can be interpreted as the Weyl factor of constant curvature metrics on the surface that we consider. We will put $\Lambda = 1$ in what follows for simplicity. Classical limit of correlator of primary operators $\exp (2a_k \phi(x)) = \exp (2\frac{a_k}{b} \varphi(x))$ depends on how their dimension scale with $b$: if $a = \eta/b \to \infty,\,b \to 0$ with $\eta$ finite (``heavy'' fields), such operators affect the equations of motion and the saddle solution, adding delta-functional terms to the RHS:
\begin{equation}
\pd \br{\pd} \varphi_{cl} =  e^{2\varphi_{cl}} - \pi \sum \limits_{k=1}^n \eta_k \delta^{(2)}(z-z_k) \label{eoms}
\end{equation}
 Leading approximation to correlation functions is then $\exp (-\frac{1}{b^2} S^{cl})$, where $S^{cl}$ is the action evaluated on the solution of (\ref{eoms}). This modification of EOMs can be interpreted as adding conical singularities (for real $0<\eta<1/2$) of angle deficit $4\pi \eta$ on the surface. For $\eta = 1/2 +i p$ these are rather macroscopic holes with geodesic boundaries of length $\sim p$. If dimensions $\Delta$ of fields stay of order $1$ when $b\to \infty$, we refer to such fields as ``light''. They do not affect the saddle point and are just evaluated on the solution, contributing the factor $\exp (2 a \varphi_{cl}/b)$ to the correlator.

After proper regularization, in simple cases we can explicitly evaluate the classical action.  For the case of three-point function it is easily verified (both for conical defects \cite{zamzam1996} and geodesic boundaries \cite{Hadasz_2004}) that the classical action coincides with the straightforward $b\to 0$ limit of the logarithm of DOZZ structure constant (\ref{liouv3pf}); assuming $\sum \limits_{i=1}^3 \eta_i>1$, the formula is
\begin{equation}
S^{(3)}_{cl}(\eta_1,\eta_2,\eta_3) = (\eta - 1) \log 2 +  F(\eta-1) + \sum \limits_{i=1}^3 F(\eta-2\eta_i) - F(0) - \sum \limits_{i=1}^3 F(2\eta_i),\,\eta \equiv \eta_1 + \eta_2 + \eta_3 \label{classc}
\end{equation}
where
\begin{equation}
F(\eta) = \psi ^{(-2)}(1-x)+\psi ^{(-2)}(x)-2 \psi ^{(-2)}\left(\frac{1}{2}\right) = \int \limits_{1/2}^\eta \log \gamma(z) dz \label{fdef}
\end{equation}
and $\psi^{(n)}$ is a polygamma function. 

An important ingredient in the study of classical Liouville theory is the $b\to 0$ limit of the BPZ equation (\ref{bpz}). Since degenerate field $V_{1,2}(z) = e^{-\varphi(z)}$ is light,  $z$-dependence factorizes out of the correlator; if the other 3 fields are heavy, we have 
\begin{equation}
   \langle V_{1,2}(z) V_{a_1} (x_1) \dots \rangle \approx_{b \to 0} \psi(z) \langle V_{a_1} (x_1) \dots \rangle \sim \psi(z) e^{-S^{cl}(\eta_1, x_1, \dots)/b^2},\,\psi(z) =  e^{-\varphi_{cl}(\eta_1,x_1,\dots)} \label{fact}
\end{equation}
We can then write the BPZ equation as follows:
\begin{equation}
[\pd^2 + t(z)] \psi(z) = 0,\,t(z) := - (\pd \varphi_{cl})^2 + \pd^2 \varphi_{cl} \label{bpzcl}
\end{equation}
In this form, it is just a consequence of Liouville EOMs and is not very helpful. However, the form of $t(z)$ can be understood even without knowing the solution, based on its singularities at $x_i$ (determined by contact terms in (\ref{eoms})),  behaviour at $\infty$ and the fact that it is holomorphic (which also follows from EOM). E.g. for correlator of 4 heavy fields $\eta_1, \dots, \eta_4$ at $x,0,1,\infty$ $t(z)$ looks like
\begin{equation}
t(z) = \frac{\delta_1}{(z-x)^2}  + \frac{\delta_2}{z^2} + \frac{\delta_3}{(z-1)^2} + \frac{x(x-1)c}{z(z-1)(z-x)} + \frac{\delta_4 - \delta_3 - \delta_2 - \delta_1}{z(z-1)},\,\delta_i = \eta_i(1-\eta_i)
\end{equation}
Parameter $c$ is called ``accessory parameter'' and cannot be determined from the requirements above (there would be more than one such parameter for  $n$-point correlators, one for each of $n-3$ complex coordinates on $n$-punctured sphere moduli space $\mathcal{M}_{0,n}$). If $\psi$ is built from the physical solution $\varphi_{cl}$ of Liouville equation, from previous factorization arguments (\ref{fact}) one expects that 
\begin{equation}
c = - \frac{\pd S^{cl}}{\pd x} \label{polyakcon}
\end{equation}
which is a so-called  Polyakov conjecture. This statement (at least for the case of the sphere) can be rigorously proven for properly regularized Liouville action (\cite{Zograf_1988}, \cite{Hadasz_2003}, \cite{Cantini_2001}). One can build from $\psi_{1,2}$ (2 independent solutions of equation (\ref{bpzcl})) a solution of Liouville equation: the function
\begin{equation}
\tilde{\varphi}_{cl} = - \log\left[ \Lambda_{ij} \psi_i(z) \br{\psi}_j(\br{z})\right]
\end{equation}
formally solves it if $\text{det }\Lambda_{ij} = -1$. For $\tilde{\varphi}_{cl}$ to be single-valued, one needs the monodromy of solutions $\psi$ to lie in the subgroup preserving the bilinear form $\Lambda$, i.e. in some real subgroup of $SL(2,\mathbb{C})$ isomorphic to either $SU(2)$ or $SU(1,1)$, depending on the sign of the curvature. This condition determines the accessory parameter $c$, for which solutions of (\ref{bpzcl}) determine $\varphi_{cl}$ that solves (\ref{eoms}).

As already mentioned, we will be interested in classical Liouville action's interpretation as a Kähler potential for certain metrics on moduli space of punctured spheres, proposed by Zograf and Takhtajan in (\cite{Zograf_1988},\cite{takhtajan2001hyperbolic}): $g^{ZT}_{i\br{j}} \sim \pd_i \br{\pd}_j S^{(cl)}$. Here $i \in 1 \dots n-3$ enumerate complex coordinates on the moduli space $\mathcal{M}_{0,n}$; a standard choice for these are $n-3$ independent cross-ratios of defects' coordinates $x_i$.
\subsection{Minimal Liouville gravity}
A CFT of total central charge 0 that consists of Liouville theory, CFT minimal model $M_{r,r'}$ and fermionic $BC$-system of central charge $-26$ (BRST ghosts) is referred to as $(r,r')$ minimal Liouville gravity (MLG):
\begin{equation}
    A_{MLG} = A_L + A_{M_{r,r'}} +\underbrace{\frac{1}{\pi}\int d^2x\,\left( C \br{\pd} B + \br{C} \pd \br{B} \right)}_{A_{ghost}} \label{mlglag}
\end{equation}
From requirement of zero total central charge it follows that Liouville parameter $b = \sqrt{r/r'}$. We will not review the properties of $BC$-CFT and minimal models in detail here; we just want to define the objects that we study in what follows. More details can be found e.g. in \cite{Zamolodchikov:2005fy}.

An important class of operators in this theory are the ``tachyons'', obtained by dressing minimal model primaries $\Phi_{m,n}$ with Liouville operators $V_a$ and ghosts $C\br{C}$ so that their total conformal dimension is $0$: in previously defined Liouville parametrization they read $W_{m,n} \equiv C \br{C} V_{m,-n} \Phi_{m,n}$. These are representatives of cohomology classes for nilpotent BRST-operator $\mathcal{Q} = C(T_L + T_M) + C \pd C B$ in this theory. Instead of adding ghosts, one can also integrate the  operators $U_{m,n} \equiv V_{m,-n} \Phi_{m,n}$ of dimension $(1,1)$ over the surface to obtained BRST-invariant objects.

Correlators of multiple tachyon operators $\int d^2x\,U_a(x)$ and $W_a(x)$ on a sphere (in fact, due to ghost number anomaly the number of $C$-ghosts in such correlator needs to be equal to three; so, three  fields are $W$ and all the others are integrated operators $\int d^2x\,U_a$) are what we consider in this paper. Such correlators do not depend on any insertion points $x_i$ and are just numbers. Moreover, in certain normalization, they turn out to simplify greatly compared to the constituents (Liouville and minimal model correlators) and become piecewise-polynomial in $m_i$ and $n_i$.

A long-standing conjecture is that these correlation numbers can be equivalently obtained from double-scaling limit of matrix models (in fact, calculations in this approach are significantly simpler than in Liouville gravity); see e.g. \cite{franc1995}, \cite{Moore:1991ir} for early studies of the problem. Identification of matrix model and LG generating functionals is complicated by the necessity to do an analytic redefinition of coupling constants, which is referred to as ``resonance transformations''; their form is most fully understood for the case of $(2,2p+1)$ MLG, corresponding to one-matrix model \cite{belzam2009}. Solidifying the connection betweeen two approaches in general case is a subject of separate line of research; in this paper we will mostly examine $4$-point correlation numbers on the sphere in $(2,2p+1)$ minimal gravity, for which calculation in continuous approach can be done and the agreement with matrix model is proven (\cite{Belavin:2005jy}, \cite{alesh2016}). In other cases, where only matrix model answers are available, we will be assuming the equivalence of two approaches and call matrix model results MLG correlation numbers.
\subsection{Some results for MLG correlators} \label{sec23} Let us order the parameters of the tachyon correlator  $\langle W_{1,k_1+1} \dots \int d^2x\,U_{1,k_n+1} \rangle$ in $(2,2p+1)$ MLG as $0 \leq k_1 \leq k_2 \leq k_3 \leq \dots \leq k_n \leq p-1$. Then the four-point correlation number, as obtained from the matrix model, reads \cite{belzam2009} 
\begin{equation}
   Z_{k_1 k_2 k_3 k_4} = -F_\theta(-2) + \sum \limits_{i=1}^4 F_\theta(k_i-1) - F_\theta(k_{12|34}) - F_\theta(k_{13|24}) - F_\theta(k_{14|23}) \label{4pfmm}
\end{equation}
where $k_{ij|lm}$ and the function $F_\theta$ are defined as
\begin{equation}
 k_{ij|lm} = \text{min} (k_i + k_j, k_l + k_m) ;\quad   F_\theta(k) = \frac{1}{2} (p-k-1)(p-k-2) \theta(p-2-k)
\end{equation}
There are corrections to this answer that are nonzero only if the so called ``fusion rules''
\begin{equation}
    \begin{cases}
    k_1 + k_2 + k_3 > k_4,\, \sum k_i \text{ is even;} \\
    k_1 + k_2 + k_3 + k_4 > 2p - 5,\,  \sum k_i \text{ is odd} 
    \end{cases} \label{fusru}
\end{equation}
are not satisfied; they nullify the correlation number in this case. The $p \to \infty$ limit of this correlator when all operators are ``heavy'' (as before, it means that parameter of the dressing Liouville fields for tachyons $\eta_{1,-k_i-1} = b a_{1,-k_i-1} = b^2 \frac{k_i+2}{2}$ or, equivalently, $\kappa_i = \frac{k_i}{p} \approx k_i b^2$ stays finite in the limit $b \to 0$) is taken using the asymptotic for function $F_\theta$:
\begin{equation} 
 F_\theta(\kappa p) \approx p^2 \cdot \frac{1}{2} (1-\kappa)^2 \theta(1-\kappa)
\end{equation}
 We also can say a bit more about the meaning of fusion rules (\ref{fusru}) in semiclassical limit. The odd-sector one becomes $\kappa_1 + \dots + \kappa_4 > 2$ and is nothing but Gauss-Bonnet theorem, which is a necessary condition for the metric on the sphere with given defects to be hyperbolic. If this inequality is not satisfied, only metric with constant positive curvature may exist; in this case, a known necessary condition for existence of such metric is the inequality $\kappa_1 + \kappa_2 + \kappa_3 > \kappa_4$, coinciding with even sector fusion rules. This inequality is referred to as Troyanov condition in the literature \cite{mazzeo2015teichmuller}.

For further reference in section \ref{sec4}, we also note the following property of these numbers: assume that sum of any two numbers $k_i + k_j,\,i \neq j$ is less than $p$ for the four-point function, the same is valid for the sum of any three numbers for five-point function and so on. Then the expression for correlation number factorizes and, up to a factor dependent only on $\sum k_i$, counts the number of conformal blocks in minimal model part of the correlator. E.g. for four-point number we have
\begin{equation}
Z_{k_1 k_2 k_3 k_4} = 
\begin{cases}
(1+k_1)(2p-3-k),\,k_{14}<k_{23} \\
(1 + \frac{k_2 + k_3 + k_1 - k_4}{2})(2p-3-k),\,k_{14} \geq k_{23} \\
\end{cases} \label{4pfconfbl}
\end{equation}
This property was noted in \cite{Artemev_2022} for four- and five-point correlation numbers based on the results of \cite{belzam2009}, \cite{tarn2011}. For higher than 5-point correlators, this behaviour can be anticipated from calculations of \cite{Fateev_2008} (although these results, relying on analytic continuation of expressions obtained from Coulomb integrals, have limited applicability in MLG, where correlators are non-analytic).
\section{Numerical calculation of moduli space volumes from CFT} \label{sec3}
In this section we describe the method for numerical calculation of ZT metrics and associated volumes. We focus on a one-parametric family of constant curvature metrics on a sphere with 4 conical defects of deficit angles $2\pi\cdot (1,1,\kappa,\kappa),\,0<\kappa<1$. Using known results in exactly solvable Liouville CFT to study classical geometry of moduli spaces was first proposed in \cite{Hadasz_2005}; our calculation is analogous to the one carried out in \cite{harrison2022liouville} (there, a case with $\kappa=1$ in our notation was studied in detail, when ZT metric coincides with the usual Weil-Petersson one for surface with punctures, or zero length geodesic boundaries) and \cite{fırat2023hyperbolic} (for torus with 1 geodesic boundary). 

The MLG result (\ref{4pfmm}) for four-point correlator suggests the following expression for the volumes in the studied case:
\begin{equation}
Z(\kappa) = 2\pi^2 \left(1 - 2 (1-\kappa)^2 + (1-2\kappa)^2 \theta(1-2\kappa) \right) \label{answer}
\end{equation}
We changed the normalization to more conventional one in geometry, where the Weil-Petersson volume $Z(1) = 2 \pi^2$; we will be more interested in parametric dependence rather than overall normalization.
\subsection{Description of the ``saddle point'' method}
We start with the known decomposition of Liouville four-point correlator
\begin{equation}
\langle V_{a_1}(0) V_{a_2}(x) V_{a_3}(1) V_{a_4}(\infty) \rangle = \int \frac{dP}{4\pi} C(a_1, a_2, \frac{Q}{2}-iP)\, C(\frac{Q}{2}+iP, a_3, a_4) \left|F_\Delta \left( 
\begin{array}{cc}
\Delta_1 & \Delta_3\\
 \Delta_2  &  \Delta_4 \\
\end{array} \right| \left. x 
\right) \right|^2 \label{decomp}
\end{equation}
Here the conformal block is normalized so that its series expansion in $x$ starts with $x^{\Delta - \Delta_1 - \Delta_2}$. If we assume that all the external and intermediate dimensions are ``heavy'' (with the parameters scaling with $b \to 0$ as before), the structure constants, as well as conformal blocks, are known to exponentiate in the $b \to 0$ limit. The integral then takes the form
\begin{equation}
\approx \int \frac{dp}{4\pi} \exp \left(- \frac{1}{b^2} \underbrace{\left[S^{(3)}_{cl} (\eta_1,\eta_2, \frac{1-2ip}{2}) + S^{(3)}_{cl} (\frac{1+2ip}{2},\eta_3,\eta_4) - 2 \text{Re }f_{\frac{1}{4}+p^2}  \left( 
\begin{array}{cc}
\delta_1 & \delta_3\\
 \delta_2  &  \delta_4 \\
\end{array} \right| \left. x 
\right)\right]}_{\equiv S^{(4)}(p,x,\br{x}) = S^{(4)}_{\text{hol}}(p,x) + S^{(4)}_{\text{a/hol}}(p,\br{x})} \right)
\end{equation}
Semiclassical structure constant $S^{(3)}_{cl}$ is given in (\ref{classc}). Then, by the usual saddle point arguments, we expect that in the semiclassical limit the integral is approximately given by \linebreak $\exp \left(-\frac{1}{b^2}S^{(4)} (p_{\text{saddle}}(x,\br{x}),x, \br{x}) \right)$ at the extremum of the expression in exponent, i.e.  $p_{\text{saddle}}$ is such that $\frac{\pd S^{(4)}}{\pd p} \mid_{p=p_{\text{saddle}}} = 0$. It is reasonable to assume (although not proven rigorously), that at least in a certain region of parameter space $S^{(4)} (p_{\text{saddle}},x, \br{x})$, computed by this method from CFT, coincides with the usual regularized classical Liouville action. It turns out that for some values of parameters $\eta$ this is not quite true (e.g. because the nontrivial real saddle disappears). We will comment on this in the following sections.

We note in passing that real $p_{\text{saddle}}$, when it exists, has a simple geometrical meaning, being proportional to the length of the (unique) simple closed geodesic separating pairs of points $(0,x)$ and $(1,\infty)$, i.e. one of the Fenchel-Nielsen  coordinates $l$. On the other hand, derivative with respect to $p$ of the holomorphic part of the action (which includes holomorphic classical conformal block and half of classical structure constants) is proportional to $i\theta$ \cite{teschner2014supersymmetric}, where $\theta$ is a conjugate twist coordinate; a saddle point condition for the integral then means that for real $p$  $\theta$ is real as well.

One can develop an expansion for $p_{\text{saddle}}$ on the boundary of moduli space (when $x \to 0$). Consider the first few terms of the expansion of $S^{(4)}$ in $p$; from explicit expressions one can see that part of $S^{(4)}(p,x,\br{x})$ coming from the structure constants is an even function of $p$, but non-analytic at $p = 0$ in such a way so we have
\begin{equation}
S^{(4)}(p, x, \br{x}) = \text{const} -A(\eta_i) |p| + B(\eta_i) p^2 + (\delta_1 + \delta_2 - \frac{1}{4} - p^2)\log (x \br{x}) + O(p^3) 
\end{equation}
Then in leading approximation the saddle point equation reads
\begin{equation}
0=\frac{\pd S^{(4)}}{\pd p} = -A + 2 |p| \log \frac{e^B}{x \br{x}} \leftrightarrow |p| = \frac{A/2}{\log \frac{e^B}{x \br{x}}} + \dots \to 0, x\to 0 \label{pasymp}
\end{equation}
It confirms that $p_{\text{saddle}}$ is small for small $x$ (note that solution exists only for $A>0$). However, all the terms that we ignored in the expansion of classical conformal blocks are ``nonperturbatively'' smaller than $p$ to any power, being proportional to powers of $q = \exp (-\#/p_{\text{saddle}})$. Ignoring them, we can first find a ``perturbative'' expansion for $p_{\text{saddle}}$ in $\frac{1}{\log (x \br{x})}$, by restoring other terms in the series for $(\pd S^{(3)}_{cl}/\pd p)$ in $p$ and solving the ``corrected'' saddle-point equation order by order. 

Instead of $x$, we will parametrize the moduli space with ``elliptic'' $q$ variable, defined as
\begin{equation}
q = \exp (i \pi \tau),\, \tau = i \frac{K(1-x)}{K(x)}
\end{equation}
where $K(x)$ is the complete elliptic integral of the first kind. To rewrite the expansion in terms of $q$ one can use that $x = 16q (1 + \dots)$. Terms denoted by $\dots$ in brackets do not matter for ``perturbative'' expansion.  We prefer ``elliptic'' parametrization because convergence of series in $q$ is generically better; also, ``non-perturbative'' corrections would be easier to construct systematically  --- one can write subleading terms in the $q$-expansion of conformal block order by order using Zamolodchikov's recursion relation \cite{Zamolodchikov1987ConformalSI} (this approach was successfully used in \cite{harrison2022liouville}). Also, we note that in principle we can expand in $\epsilon = (\log a(\eta_i)/q\br{q})^{-1}$ for any function $a(\eta)$; for a specific choice $a = e^B/2^8$ quadratic in $p$ terms in the expansion have the most simple form. 

\begin{figure}[h] 
    \centering
    \includegraphics[height=0.2\textheight]{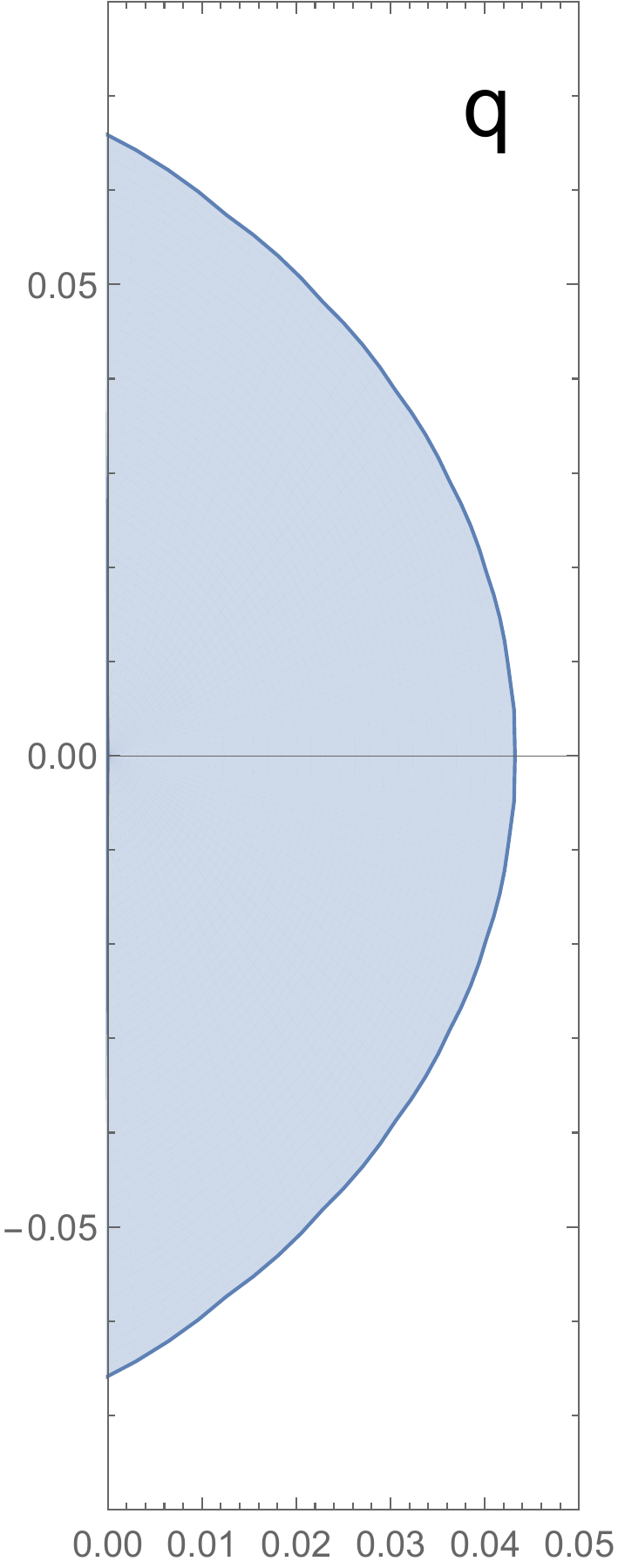}
     \includegraphics[height=0.2\textheight]{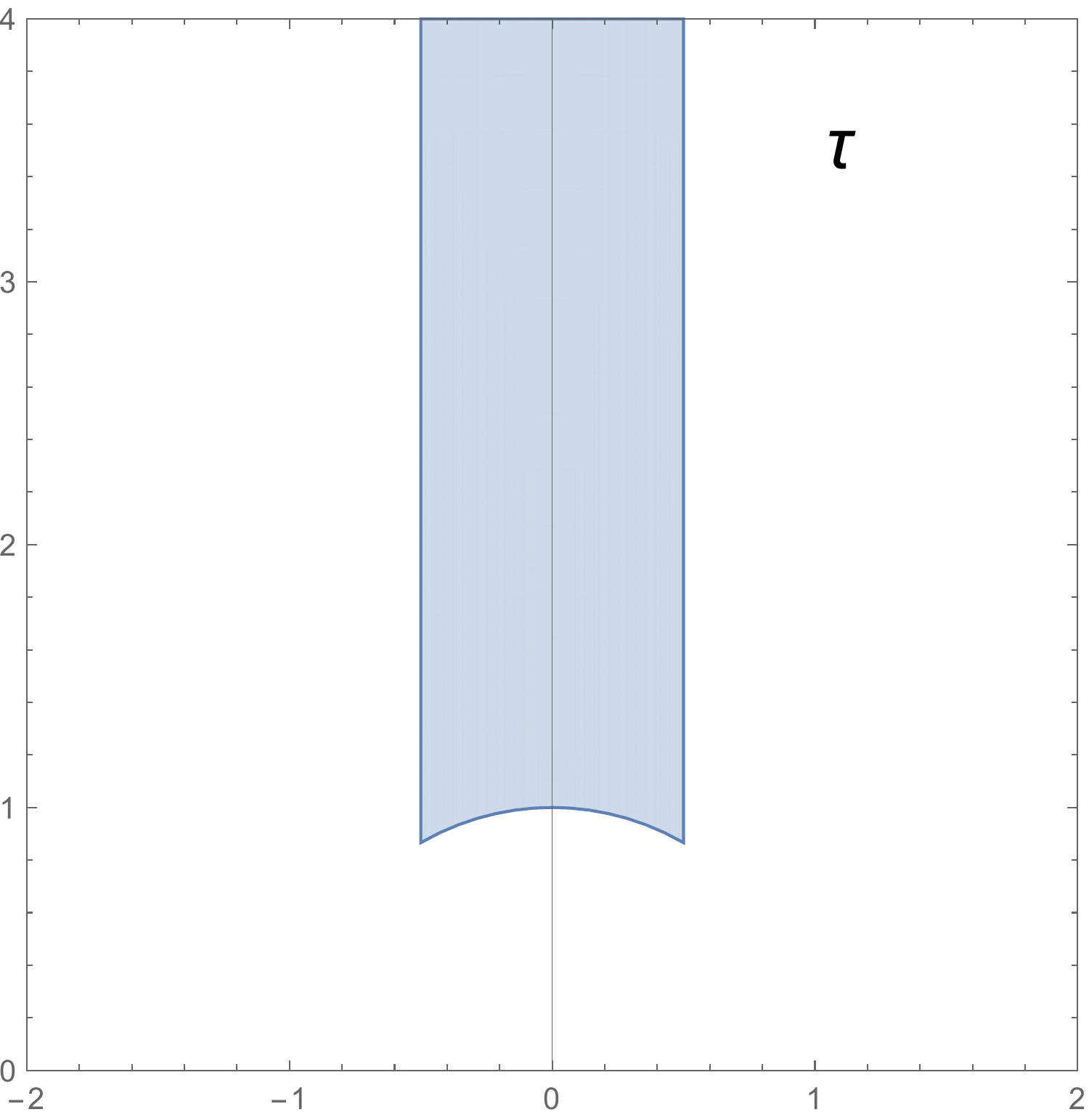}
    \includegraphics[height=0.2\textheight]{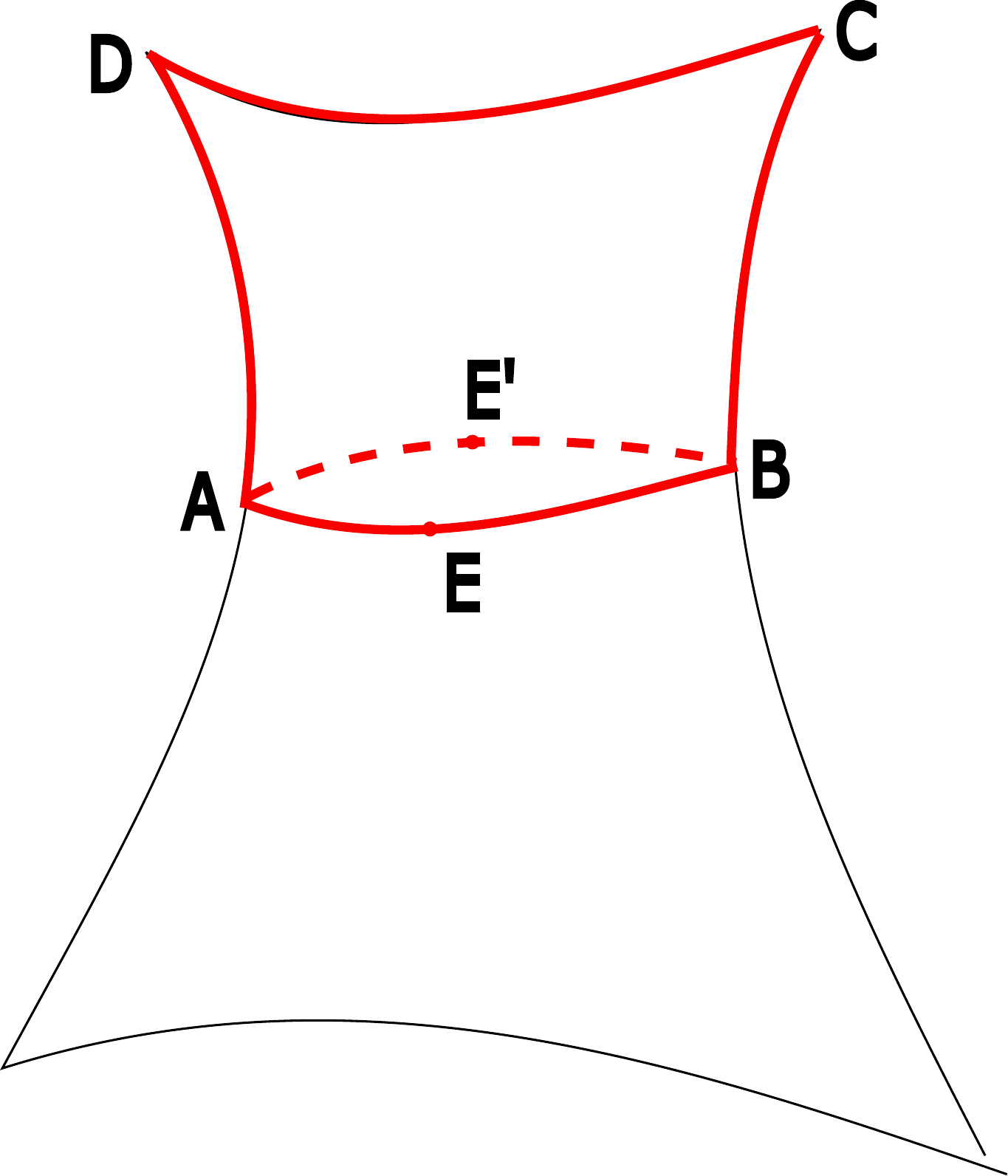}
    \caption{1 and 2: fundamental domain $\mathcal{F}$ in $q$ and $\tau$ coordinates. 3: Hyperbolic surface with 4 conical defects and the separating geodesic}
    \label{fig:msandgeodesics}
\end{figure}

After the series for $p_{\text{saddle}}$ is found, we can compute the ``classical action'' $S^{(4)}$ and the associated Kähler metric on the moduli space: $g_{q\br{q}} = -4\pi \pd_q \pd_{\br{q}} S^{(4)}(q,\br{q})$ (as a series expansion in $\epsilon$). Up to this step, obtaining analytic expressions is possible; the only thing left to compute the volumes is to integrate $\sqrt{\text{det }g}$  over the moduli space $\mathcal{M}_{0,4}$. The discrete group $S_3 = S_4/(S_2 \times S_2)$ of order 6 permutes the conical defects and, nontrivially acting on cross-ratio $x$, acts on $\mathcal{M}_{0,n}$. A convenient choice of fundamental domain of this action is
\begin{equation}
\mathcal{F} = \lbrace q = r e^{i\psi}\mid -\pi/2 \leq \psi\leq\pi/2,\,0\leq r\leq \exp (-\sqrt{\pi^2 -\psi^2}) \rbrace
\end{equation}
It is depicted on Fig. \ref{fig:msandgeodesics}, but it is easier to understand how this domain looks in $\tau$ coordinate, where it is just the usual fundamental domain for $SL(2,\mathbb{Z})$ action on the upper half-plane. Integration over this domain of the terms in the series for $g_{q \br{q}}$ can be carried out numerically.

\subsection{Example of application and problems}
Let us now specify previous results for an example that we introduced in the beginning of this section: the four-point function $\langle V_{\kappa/2b}(0) V_{\kappa/2b}(x) V_{1/2b} (1) V_{1/2b}(\infty) \rangle$. The moduli space $\mathcal{M}_{0,4}$ is decomposed into a fundamental domain $\mathcal{F}$ and $5$ its images under the action of the aforementioned $S^3$ group; in 2 of these 6 domains, $V_{\kappa/2b}(x)$ is closer to the operator $V_{\kappa/2b}(0)$ and should be fused with it, while in the other 4 it should be fused with $V_{1/2b}$. To calculate the integral over the whole moduli space, in each domain we should use decomposition like the formula (\ref{decomp}) in a corresponding channel. 

First, we need to understand when the real saddle point, that we need to evaluate the integral, exists. Using the formula (\ref{classc}) for classical structure constants, we get that for $(1/2, \kappa/2)$ fusion channel for all $0<\kappa<1$ the coefficient $A$ in (\ref{pasymp}) is equal to $2\pi>0$. Thus, the real saddle exists and in the corresponding domains the calculation procedure is reliable. However, for the $(\kappa/2, \kappa/2)$ channel this is not the case --- the coefficient $A$ is equal to $2\pi$ for $\kappa>1/2$ and otherwise is \emph{zero}. For more general values of parameters of the correlator, real saddle in the integral (\ref{decomp}) would disappear if the parameters $\eta_1, \eta_2$ of fused fields are such that $\eta_1 + \eta_2<1/2$. 

The formal reason for such behaviour is the following --- functions $F(\eta_1 + \eta_2 - 1/2 - ip)$ and $F(\eta_1 + \eta_2 - 1/2 + i p)$, entering the classical structure constants (\ref{classc}) and defined by integral representation (\ref{fdef}), develop an additional linear in $p$ contribution to the Maclaurin series in $p$ when $\text{Re }(\eta_1 + \eta_2 - 1/2)<0$. Indeed, the integrand $\log \gamma(z)$ in (\ref{fdef}) obtains an additional imaginary part $\pm i \pi$; from this the proposed behaviour of $F(\eta)$ follows. These additional linear terms from two $F$-functions above then cancel the term $-2\pi |p|$ that series for $S^{(4)}$ had when $\kappa>1/2$.

We were a little bit inaccurate in the arguments of the previous paragraph, using naive analytic continuation for classical structure constants (\ref{classc}) in the domain $\eta_1 + \eta_2 < 1/2$. In fact, as noted before in \cite{Harlow_2011}, in this case classical limit of the three-point function changes less trivially: if real part of the argument of some $F$-function in (\ref{classc}) is $-1<\eta<0$, to get the asymptotic coinciding with the logarithm of DOZZ formula one needs to replace
\begin{equation}
\exp \left(- \frac{1}{b^2} F(\eta) \right) \Rightarrow \frac{1}{2\sinh \frac{i \pi \eta}{b^2}} \exp \left(- \frac{1}{b^2} F(\eta+1) \right)
\end{equation}
For $\text{Im }\eta = \pm p$ the denominator can be expanded in series in $\exp(- \pi |p|/b^2)$. This expansion can be interpreted as sum over certain complex saddle points in the functional integral. Performing the replacement above for 2 $F$-functions with negative real part of their argument and proceeding with the expansion, it is easy to see that there is no nontrivial real saddle for integral over $p$ in any of the terms in the obtained series.
 
Recalling that $p_{\text{saddle}}$ is proportional to the length of separating geodesic, we can understand the geometric meaning of why the saddle point disappears: on a hyperbolic surface such geodesic does not exist if conical defects are not sharp enough. Indeed, consider Fig. \ref{fig:msandgeodesics}; suppose that geodesic $AEBE'A$ that separates two pairs of points exist. Then, together with the parts of geodesics that connect the defects, we obtain 2 hyperbolic tetragons $AEBCD$ and $AE'BCD$. The sum of all their angles is equal to $2\pi (2-2\eta_1 - 2\eta_2) + 2\pi$ and by Gauss-Bonnet theorem should be less than $4\pi$, which leads necessarily to $\eta_1 + \eta_2>1/2$.

Summarising this discussion, we see that the procedure that we use is only reliable for $\kappa>1/2$. Thus, we will only restrict to such  values of the parameter in the next section.
\subsection{Results of the calculation}
\begin{figure}[h] 
    \centering
    \includegraphics[width=0.4\textwidth]{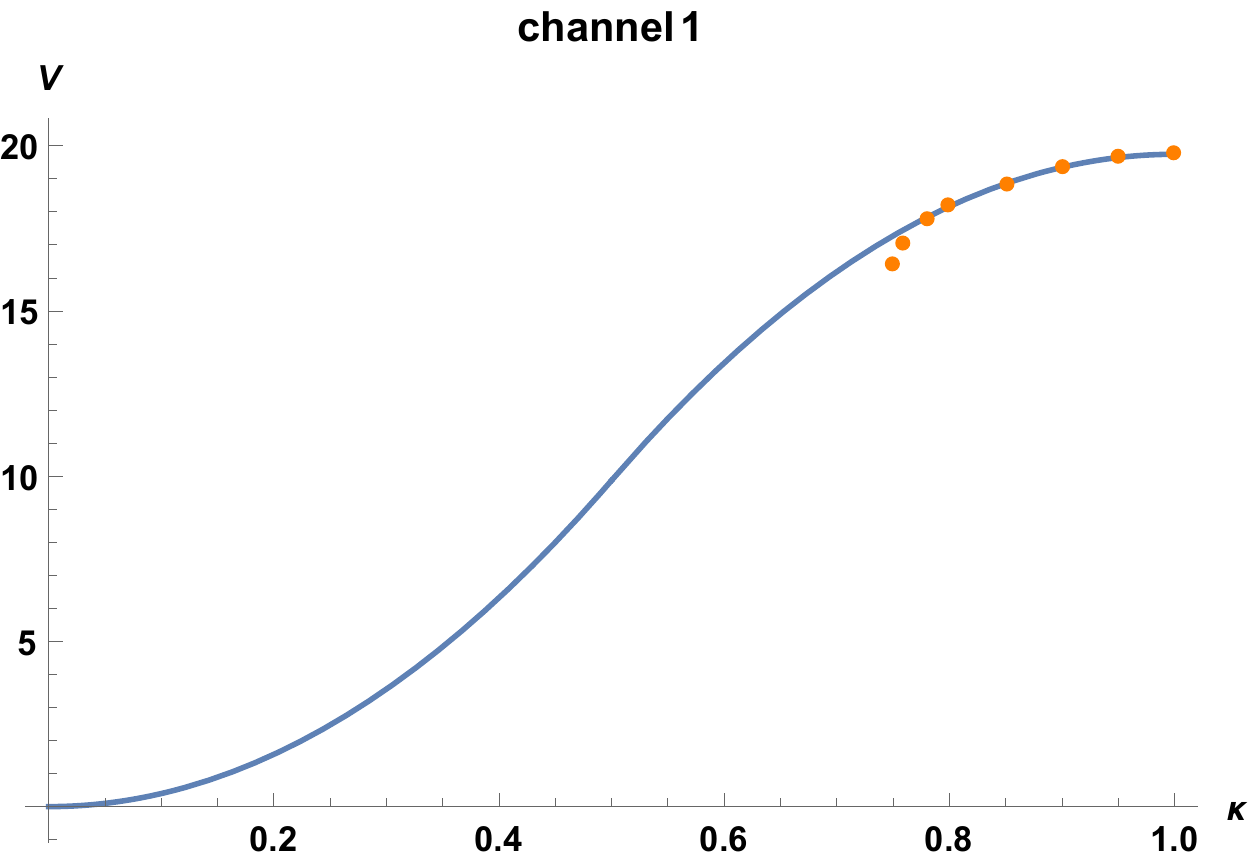}
    \includegraphics[width=0.4\textwidth]{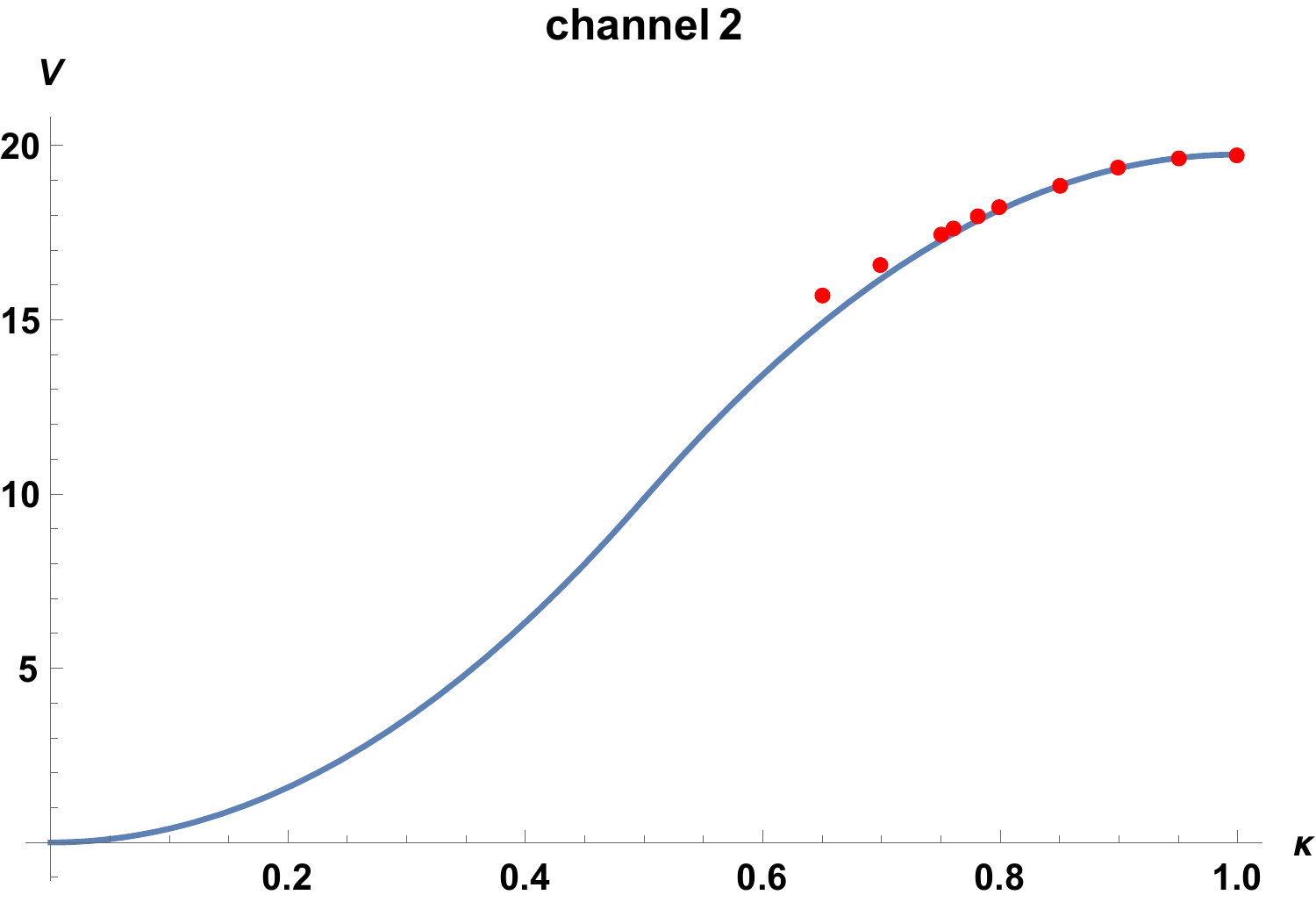}
    \caption{$\kappa$-dependence of moduli space volume contributions from 2 channels (times 6). At $\kappa = 0.7$  agreement is already bad.}
    \label{fig:plotsbad}
\end{figure}

For brevity we will call $(\kappa/2, \kappa/2)$ and $(1/2, \kappa/2)$ channels ``channel 1'' and ``channel 2'' respectively. We introduce the ``perturbative'' expansion parameter $\epsilon$ as
\begin{equation}
\frac{1}{\epsilon} = \log \frac{1}{x \br{x}} - f(\delta)
\end{equation}
where $\delta \equiv \kappa - \frac{1}{2}$ and $f$ is some function of $\delta$, independent of $x$. Then, the saddle point equation in ``perturbative'' approximation becomes
\begin{equation}
\frac{\pd S^{(4)}}{\pd p} = \underbrace{\frac{\pd}{\pd p} \left(S^{(3)}_{cl} (\eta_1,\eta_2, \frac{1}{2}-ip) + S^{(3)}_{cl} (\frac{1}{2}+ip,\eta_3,\eta_4) \right)}_{= -2\pi |p| + \dots} + 2p \left(\frac{1}{\epsilon} + f(\eta)\right) \mid_{p=p_{\text{saddle}}} = 0
\end{equation}
and can be solved order by order in $\epsilon$, putting $p_{\text{saddle}} = \sum \limits_{n=1}^{n_{max}} p_n \epsilon^n$. For the following choice of $f$ for channels $1$ and $2$ respectively
\begin{equation}
f_1 = \psi ^{(0)}(1-\delta )+\psi ^{(0)}(\delta )+8 \gamma +6 \psi ^{(0)}\left(\frac{1}{2}\right) \text{ and } f_2 = 4 \psi ^{(0)}\left(\frac{1}{4} (2 \delta +1)\right)+4 \psi ^{(0)}\left(\frac{1}{4} (3-2 \delta )\right)+8 \gamma \label{fchoice}
\end{equation}
expansion coefficients look the simplest; e.g. up to 6th order in $\epsilon$ the saddle point momentum is
\begin{align}
&p_\text{saddle1} = \pi  \epsilon -\frac{1}{6} \epsilon ^4 \pi ^3 \left(\psi ^{(2)}(1-\delta )+\psi ^{(2)}(\delta )+6 \psi ^{(2)}\left(\frac{1}{2}\right)-32 \psi ^{(2)}(1)\right)+ \nonumber \\
&+ \frac{1}{120} \pi ^5 \epsilon ^6 \left(\psi ^{(4)}(1-\delta )+\psi ^{(4)}(\delta )+6 \psi ^{(4)}\left(\frac{1}{2}\right)-128 \psi ^{(4)}(1)\right)+ \dots
\end{align}
in channel 1 and
\begin{align}
&p_\text{saddle2} = \pi  \epsilon +\frac{2}{3} \epsilon ^4 \pi^3 \left(-\psi ^{(2)}\left(\frac{1}{4} (3-2 \delta )\right)- \psi ^{(2)}\left(\frac{1}{4} (2 \delta +1)\right)+8  \psi ^{(2)}(1)\right) + \nonumber \\
&+ \frac{1}{30} \epsilon ^6 \pi^5 \left(\psi ^{(4)}\left(\frac{1}{4} (3-2 \delta )\right) +  \psi ^{(4)}\left(\frac{1}{4} (2 \delta +1)\right)-32 \psi ^{(4)}(1)\right)+ \dots
\end{align}
in channel 2. We can now substitute that in the action/Kähler potential, obtaining the series in $\epsilon$, differentiate it termwise using
\begin{equation}
\pd_q \pd_{\br{q}} \epsilon^{-n} = \frac{n(n+1)}{q \br{q}}  \epsilon^{-n-2}
\end{equation}
and obtain the series expansion for the metric in 2 channels (again, for simplicity we write the formulas for the choice of $f$ as in (\ref{fchoice}))
\begin{equation}
g_{1} = \frac{8\pi^3 \epsilon^3}{q \br{q}} \left(1 -10 \pi^2 \left(\frac{\psi ^{(2)}(1-\delta )+ 6 \psi ^{(2)}\left(\frac{1}{2}\right)+  \psi ^{(2)}(\delta )}{12}+\frac{8 \zeta (3) - 4 \psi ^{(2)}(1)}{3} \right) \epsilon^3  + \dots \right)
\end{equation}
\begin{equation}
g_{2} = \frac{8\pi^3 \epsilon^3}{q \br{q}}  \left(1 - 10 \pi^2 \left( \frac{\psi ^{(2)}\left(\frac{2 \delta +1}{4}\right)+  \psi ^{(2)}\left(\frac{3-2 \delta}{4} \right)}{3} +\frac{8 \zeta (3) - 4  \psi ^{(2)}(1)}{3} \right) \epsilon^3 + \dots \right)
\end{equation}
Then, we integrate each term over the fundamental domain $\mathcal{F}$. Radial integration can be performed exactly
\begin{equation}
\int \limits_0^{\exp (- \sqrt{\pi^2 - \psi^2})} \frac{dr}{r} (\text{const} - 2 \log r)^{-n} = \frac{\left(\text{const} +2 \sqrt{\pi ^2-\psi ^2}\right)^{1-n}}{2 (n-1)}
\end{equation}
and integration over the angle $\psi$ --- only numerically. Expanding to 25th order in $\epsilon$, integrating over the fundamental domain in two channels, we obtain the following plots (Fig. \ref{fig:plotsbad}).

When $\kappa$ is close enough to 1, agreement is very good. However, expansion coefficients for the metric start growing too quickly if we change $\kappa$. The final numeric sum doesn't seem to converge when truncated to the order that we consider (the highest one we considered was 50th order in $\epsilon$). E.g. on Fig. \ref{fig:tablebad} we have the results of integration of different terms in the series for the metricin channel 1 for $\delta = 0.25$: 
  \begin{figure}[h] 
    \centering
    \includegraphics[width=\textwidth]{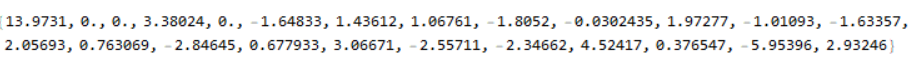}
    \caption{Contributions to moduli space volume from channel $1$ from integrating different terms in the series for the metric, starting with $1/(q \br{q} \log^3 |q|)$. Absence of convergence can be seen.}
    \label{fig:tablebad}
\end{figure}

\begin{figure}[h] 
    \centering
    \includegraphics[width=0.4\textwidth]{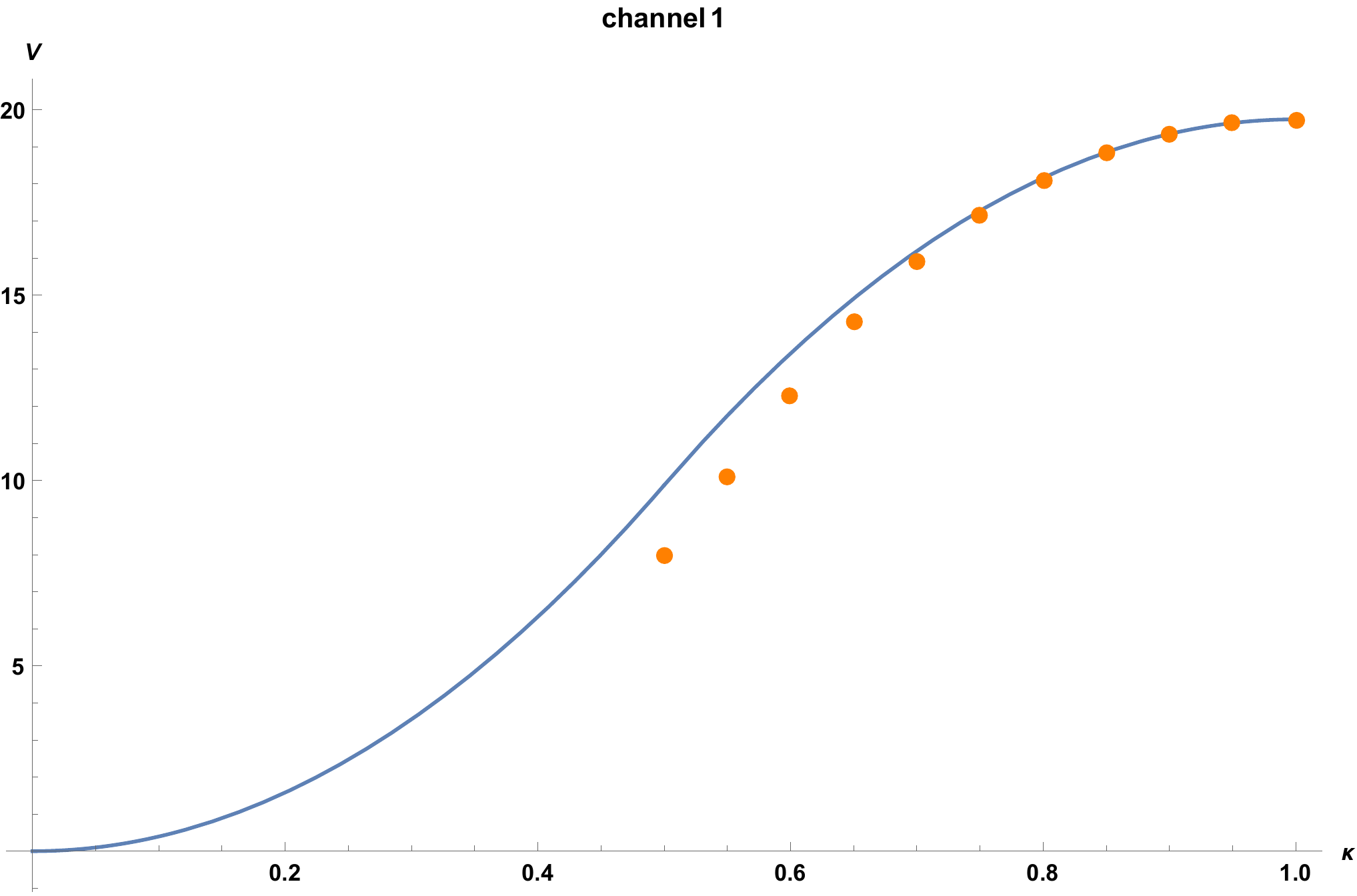}
    \includegraphics[width=0.4\textwidth]{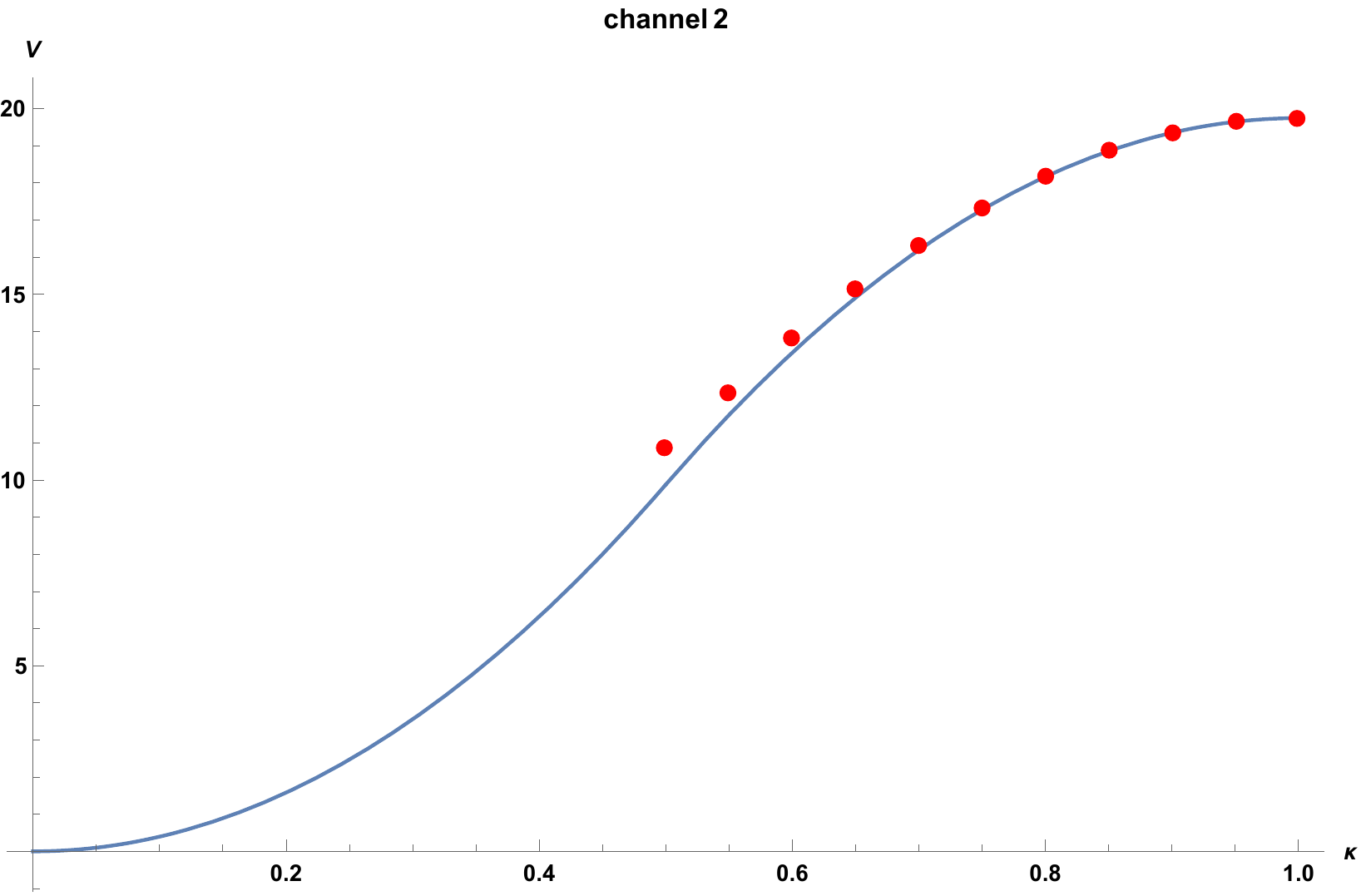}
    \caption{$\kappa$-dependence of contributions to moduli space volume from 2 channels (normalized to agree with the full answer at $\kappa = 1$) with tuned expansion parameter}
    \label{fig:plotsgood}
\end{figure}

However, it turns out that this problem can be tamed by tuning $f(\delta)$ and the expansion parameter $\epsilon$. In Table \ref{tablegood} and Figure \ref{fig:plotsgood} for each data point we fitted $f(\delta)$ to obtain optimal convergence (also, for some points it was necessary to extend the expansion from 25th to 40th order in $\epsilon$).
\begin{figure}
\centering
\small
\begin{tabular}{|c|c c c c c c c c c c c|}
\hline
     $\kappa$ & 1 & 0.95 & 0.9 & 0.85 & 0.8 & 0.75&  0.7 & 0.65 & 0.6 & 0.55 & 0.5 \\
     \hline
     \text{ch. 1} & 19.737 & 19.639 & 19.344 & 18.856 & 18.181 & 17.326 & 16.303 & 15.125 & 13.810 & 12.381 & 10.865 \\
     \text{ch. 2} & 19.737 & 19.639 & 19.340 & 18.835 & 18.115 & 17.137 & 15.913 & 14.265 & 12.287 & 10.082 & 7.990 \\     
     \text{sum} & 19.737 & 19.639 & 19.343 & 18.850 & 18.159 & 17.263 & 16.173 & 14.838 & 13.302 & 11.615 & 9.907 \\      
     (\ref{answer}) & 19.739 & 19.641 & 19.344 & 18.851 & 18.160 & 17.272 & 16.186 & 14.903 & 13.423 & 11.745 & 9.870 \\       
     \hline
\end{tabular}
    \caption{Numeric data for contributions from different channels (data points for channel 1 and channel 2 are normalized to agree with the full answer at $\kappa = 1$)} 
    \label{tablegood}
\end{figure}

\begin{figure}[h] 
    \centering
    \includegraphics[width=0.5\textwidth]{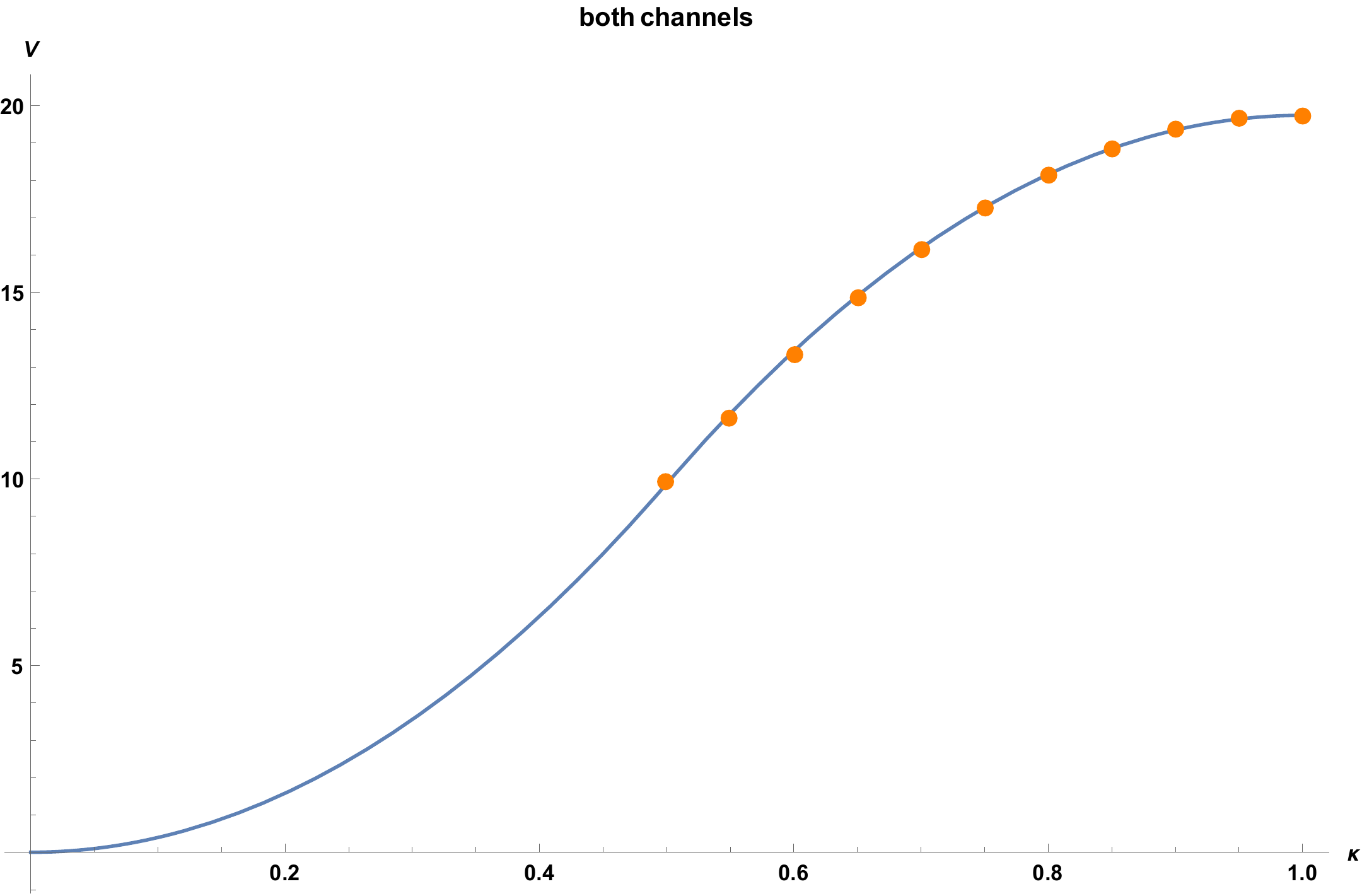}
    \caption{Sum of contributions from both channels for tuned parameter $f$}
    \label{fig:plotgoodfinal}
\end{figure}

Contributions in 2 different channels start to look different from each other and the analytic prediction (\ref{answer}) when $\kappa$ is small enough. However, the appropriate sum over all 6 images of fundamental domain is in quite good agreement (accuracy $\sim 1 \%$) with the analytic answer for all considered values of parameter $\kappa$ (see Fig. \ref{fig:plotgoodfinal}).
\section{Classical action for ``perturbatively heavy'' insertions} \label{sec4}
\subsection{Motivation: ``light limit''} \label{lightlim}
The simplification of MLG answers when $p>k_i + k_j$ mentioned near (\ref{4pfconfbl}) is suggestive of how the moduli space metric behaves for these values of parameters. Consider e.g. the sphere 4-point function in the case when $k_1+k_4<k_2+k_3$, which means that the number of conformal blocks is maximal and equal to $(1+k_1)$ (as before we assume $k_1 \leq k_2 \leq k_3 \leq k_4\leq p-1$), MLG correlator and its semiclassical limit are (\ref{4pfconfbl})
\begin{align}
    &Z_{k_1 k_2 k_3 k_4} = (1+k_1)(2p-3-k_1-k_2-k_3-k_4) \approx p^2 \kappa_1 (2 - \sum \kappa_i),\,p \to \infty; \nonumber \\
    & \kappa_i \equiv \frac{k_i}{p} \approx 2b a^L_{1,-k_i-1} = 2\eta_i \label{z4}
\end{align}
For general $n$-point correlator on the sphere under the same conditions apparently we would have (see the discussion at the end of \ref{sec23})
\begin{equation}
    Z_{k_1 k_2 k_3 \dots k_n} \approx p^{2n-6} \kappa_1 \kappa_2 \dots \kappa_{n-3} (2 - \sum \kappa_i)^{n-3},\,p \to \infty \label{zn}
\end{equation} 
Before semiclassical limit, formula is a bit more complicated: the factor other than the number of conformal blocks is most easily expressed via number of screenings, see \cite{Fateev_2008}. Note that this factor is just the RHS of Gauss-Bonnet theorem 
\begin{equation}
 \text{Area}[g] \sim 2-\sum \kappa_i
\end{equation}
to the power $(n-3)$. This LHS of this theorem, on the other hand, is just the volume of the ``physical'' sphere where our constant curvature metric is defined. From the complex geometry point of view, moduli space  is just a symmetrized product of $n-3$ spheres (up to blowups at singular points). Then, as we see that moduli space volume (\ref{zn}) is proportional to to $(\text{Area}[g])^{n-3}$, we can conjecture that (up to a coefficient) in this limit the moduli space metric reduces to metric on the product space of $n-3$ spheres with ``physical`` metric $g$. 

This property is easily understood from ZT proposal in the ``light'' limit for Liouville correlators. In this limit dimensions of fields $\Delta_i = a_i (Q-a_i) = O(1),\,b\to 0$ for $i=1 \dots n-3$; so only 3 heavy fields $\kappa_{n-2}, \kappa_{n-1},\,\kappa_n$ affect the classical solution of Liouville equation $\varphi_{cl}$. Nontrivial $x_i$ dependence factorizes from the correlator as $\exp \left( \frac{1}{b} \sum a_i \varphi_{cl}(x_i) \right)$. 
Considering the argument of the exponent as a Kähler potential, the corresponding metric $g^{ZT}_{i\br{j}} = \pd_i \br{\pd}_j S$ becomes block-diagonal: $g^{ZT}_{i\br{j}}(x_k) = \delta_{ij} a_i \pd \br{\pd} \phi(x_i)$. Due to Liouville equation $\pd \br{\pd} \varphi = e^{2\varphi}$ this indeed reduces to the metric $g = e^{2\varphi} dz\,d\br{z}$ on the physical sphere up to coefficient. The volume associated with this metric is
\begin{align}
&\int \prod \limits_{i=1}^{n-3}d^2x_i\, \sqrt{\text{det }g^{ZT}} = \prod \limits_{i=1}^{n-3}\left(\int d^2x_i\, a_i e^{2 \varphi(x_i)} \right)  = \prod \limits_{i=1}^{n-3} a_i \left(\int d^2x_i\, \sqrt{\text{det }g} \right) = \nonumber \\
&=(2-\kappa_{n-2}-\kappa_{n-1}-\kappa_{n})^{n-3} \prod \limits_{i=1}^{n-3}a _i
\end{align}
To leading order in $a_i$ this is consistent with MLG answer (\ref{zn}). It is not precisely what we want, since the asymptotic of MLG answer that we take implies that all fields should be heavy, with dimension $\Delta_i = O(1/b^2)$. However, we expect that the arguments about factorization should also be valid in the so-called ``perturbatively heavy'' limit --- when after taking the limit $b \to 0, a_i = \eta_i/b$, $\eta_i$ finite, we take the limit $\eta_i \to 0$ (or expand in series in $\eta_i$). This limit for conformal blocks is particularly meaningful and was extensively studied in the AdS${_3}$/CFT${_2}$ context (see e.g. \cite{Alkalaev_2020}).

A priori it is not obvious that ``light'' and ``perturbatively-heavy'' limit agree with each other; in the following sections we will confirm these expectations by an explicit calculation.
\subsection{Monodromy method for calculating the classical action}
We remind that for the four-point function $\langle V_{1,2}(z) V_{a_1}(x) V_{a_2}(0) V_{a_3}(1)  V_{a_4}(\infty) \rangle$ classical BPZ equation (\ref{bpzcl}) reads
\begin{equation}
[\pd_z^2 + t(z)] \psi = 0,\,t(z) = \frac{\delta_1}{(z-x)^2}  + \frac{\delta_2}{z^2} + \frac{\delta_3}{(z-1)^2} + \frac{x(x-1)c}{z(z-1)(z-x)} + \frac{\delta_4 - \delta_3 - \delta_2 - \delta_1}{z(z-1)}
\end{equation}
One can use this equation to study both the ``holomorphic'' problem of determining classical conformal blocks and ``mixed'' one for classical Liouville action. In both cases we need to vary an accessory parameter $c(x)$ to realise certain conditions on monodromy of the system of solutions. For conformal block traces of monodromy matrix are determined by intermediate dimensions of the block that we want to study; for classical Liouville action, as mentioned before, the conditions are that the monodromy group is isomorphic to a real form of $SL(2,\mathbb{C})$ --- either $SU(1,1)$ or $SU(2)$. After finding $c(x)$, either conformal block or classical action (more precisely, its coordinate-dependent part) can be found from relation (\ref{polyakcon}).

Expansion in $\eta_i$ for the solution of the equation above can be developed as follows. We separate
\begin{equation}
t(z) =t^{(0)}(z) + t^{(1)}(z),\,
\end{equation}
with $t^{(1)}$ containing contributions proportional to perturbatively small dimensions and the accessory parameter, and look for the solution as a series  $\psi = \psi^{(0)} + \psi^{(1)} + \dots$. Equating order by order, we get the following chain of equations
\begin{equation}
    \begin{cases}
    [\pd_z^2 + t^{(0)}]\psi^{(0)} = 0\\
    [\pd_z^2 + t^{(0)}]\psi^{(1)} = -t^{(1)}\psi^{(0)}\\
    [\pd_z^2 + t^{(0)}]\psi^{(2)} = -t^{(2)}\psi^{(0)} - t^{(1)} \psi^{(1)}\\
    \dots \\
    \end{cases}
\end{equation}
Solutions of the zero-order equation $\psi^{(0)}_{\pm}$ are assumed to be known (the cases we consider are with $2$ and $3$ heavy operators, where they can be found explicitly). Then, corrections $\psi^{(i)}$ can be found using the method of variation of parameters. At first order we have
\begin{equation}
\psi^{(1)}_\pm = \frac{1}{W} \left( \psi^{(0)}_+ \int \limits^z \psi^{(0)}_-  t^{(1)} \psi^{(0)}_\pm  -  \psi^{(0)}_- \int \limits^z \psi^{(0)}_+ t^{(1)}\psi^{(0)}_\pm \right)
\end{equation}
($W$ is the wronskian of solutions $\psi_\pm^{(0)}$). Monodromy matrix to first order is determined by the integrals $\oint \psi_{\mp}^{(0)} T^{(1)} \psi_{\pm}^{(0)}$ around singular points, i.e. the residues of the integrands. Specifically, monodromy matrix for contour $\gamma$
\begin{equation}
M^{(1)}_\gamma = (1_{2\times 2} + I) M^{(0)}_{\gamma},\,I = \frac{1}{W} 
\begin{pmatrix}
\oint \limits_\gamma \psi_-^{(0)} t^{(1)} \psi_+^{(0)} & -\oint  \limits_\gamma \psi_+^{(0)} t^{(1)} \psi_+^{(0)} \\
 \oint  \limits_\gamma\psi_-^{(0)} t^{(1)} \psi_-^{(0)}   & -\oint  \limits_\gamma \psi_+^{(0)} t^{(1)} \psi_-^{(0)} \\
\end{pmatrix}    
\end{equation}
Extension of this procedure to higher orders apparently meets some difficulties \cite{trufpopova}. The reason is that one can not simply close the contour of integration and calculate the residues, since the integrand will not be single-valued in the vicinity of singular points. In the following sections we restrict to first order in parameters of perturbatively heavy fields, although in fact expressions for the volumes we are looking for should be exact in some order in such perturbation theory, because volumes are polynomial in $\eta_i$. It might be interesting to find a way to see it explicitly.
\subsection{4 defects: example wih 2 heavy operators} \label{sec43}
The simplest case with 2 heavy ($\eta_3 = \eta_4 = \eta_h$; we need to have heavy dimensions equal for classical solution of Liouville equation to exist) and 2 perturbatively heavy ($\eta_1 = \eta_2 = \eta_l$) operators was studied in \cite{Balasubramanian_2017} where the following coordinate dependence of classical action was established:
\begin{equation}
S_{cl}(x) = S = 4 \eta_L \log (1 + |1-x|^{1-2\eta_h}) + F(x) + F(\br{x}) + O(\eta_L)^2
\end{equation}
Metric volume form calculated from this action can be integrated over $d^2x$, yielding $\sim \eta_l (1-2\eta_h) \sim \kappa_l (2 - 2 \kappa_h)$ as the volume. This is consistent with (\ref{z4}) to first order in $\eta_l$. 

Here we slightly generalize the analysis of \cite{Balasubramanian_2017} to the case of 2 different light operators $\eta_1 = \eta_l,\,\eta_2 = \eta_l \cdot \sigma$. Decomposition of the energy-momentum tensor $t(z)$ in this case is
\begin{equation}
t^{(0)}(z) =  \frac{\delta_3}{(z-1)^2},\,t^{(1)}(z) = \frac{\eta_l}{(z-x)^2}  + \frac{\sigma \eta_l}{z^2} + \frac{x(x-1)c}{z(z-1)(z-x)} - \frac{\eta_l(1+\sigma)}{z(z-1)}
\end{equation}
and the monodromy matrices for basis of zero-order solutions
\begin{equation}
\tilde{\psi}_\pm (z) = (1-z)^{\frac{1 \pm \alpha}{2}}
\end{equation}
are ($C \equiv c/\eta_l$ and $\alpha \equiv 1-2\eta_h$)
\begin{equation}
M_{\gamma_0} = 1_{2 \times 2} + \frac{2\pi i \eta_l}{\alpha} 
\begin{pmatrix}
C(1-x) + \sigma & -C(1-x) + \alpha - \sigma\\
C(1-x) + \alpha+ \sigma & - C(1-x) - \sigma \\
 \end{pmatrix}
\end{equation}
\begin{equation}
M_{\gamma_x} = 1_{2 \times 2} + \frac{2\pi i \eta_l}{\alpha} 
\begin{pmatrix}
-C(1-x) - \sigma & (C(1-x) + \sigma(1+ \alpha))(1-x)^\alpha\\
-(C(1-x) + \sigma(1-\alpha)) (1-x)^{-\alpha} & C(1-x) + \sigma \\
 \end{pmatrix}
\end{equation}
Unitarity condition to first order in $\eta_L$ for $0$ and $x$ reads
\begin{equation}
J_0 \delta M_\gamma = (\delta M_\gamma^{-1})^\dagger J_0 = - \delta M_\gamma^\dagger J_0 \label{su2cond}
\end{equation}
where $\delta M_\gamma$ is a linear in $\eta_l$ part of monodromy matrix,  $J = B^\dagger B$ and $B$ is the $SL(2,\mathbb{R})$ matrix of change of basis from $\tilde{\psi}_{\pm}$ to the basis with $SU(2)$-monodromy around all punctures. From unitarity condition at 0th order we should have $J_0 = \text{diag}(a, 1/a)$; then, matrix equation (\ref{su2cond}) reduces to 2 equations for $C$
\begin{equation}
\begin{cases}
    C(1-x) = \br{C} (1-\br{x}) \\
    \frac{C(1-x) + \sigma(1-\alpha)}{C(1-x) + \sigma(1+\alpha)} \frac{1}{|1-x|^{2\alpha}} = \frac{C(1-x) + \alpha + \sigma}{C(1-x) + \sigma-\alpha} \\
\end{cases}
\end{equation}
This equations are solved by 
\begin{equation}
C = \frac{1}{1-x} \left(\frac{\alpha  \left(\sqrt{\sigma ^2 (\zeta +1)^2-2 \sigma  ((\zeta -6) \zeta +1)+(\zeta +1)^2}-(\sigma +1) (\zeta +1)\right)}{2 (\zeta -1)}-\sigma \right),
\end{equation}
where $\zeta \equiv |1-x|^{2\alpha} $. We can differentiate it over $\br{x}$ and obtain the metric; integrating this bulky expression over $x$, however, is difficult. Instead, consider how the prefactor behaves at $x \to 1$ and $x \to \infty$:
\begin{equation}
C \approx \frac{1}{1-x} \left(\alpha \cdot \frac{\sigma+1 - \sqrt{(\sigma-1)^2}}{2} -\sigma\right),\,x \to 1
\end{equation}
\begin{equation}
C \approx -\frac{1}{x} \left(\alpha \cdot \frac{\sqrt{(\sigma-1)^2}-(\sigma+1)}{2} -\sigma\right),\,x \to \infty
\end{equation}
Since metric coefficients (which in this case are the same as coefficients of the volume form) are given by $g_{x \br{x}} \sim \br{\pd} c = \pd \br{c}$, integral over $d^2x$ reduces to sum of boundary terms, which are residues of $C$ and $\br{C}$ at $x=1$ and $\infty$ (with an appropriate sign):
\begin{equation}
V \sim \eta_L \cdot \alpha \cdot \left(\sigma+1 - \sqrt{(\sigma-1)}^2 \right)  = \eta_L (2-4\eta_h) \cdot
\begin{cases}
 1,\,\sigma>1\\
 \sigma,\,\sigma<1 \\
\end{cases}
\end{equation}
This is, again, consistent with the answer (\ref{z4}) from MLG: for the case that we consider, number of conformal blocks is defined by the smallest of 2 numbers $\eta_1,\eta_2$, which depends on whether $\sigma$ is greater or smaller than 1. 
\subsection{4 defects: example with 3 heavy operators}
Now let's try to use the results of \cite{Alkalaev_2019} for the case of 1 light operator with parameter $\epsilon_2$ at point $z$ and three heavy operators with parameters $\kappa_2/2,\kappa_1/2,\kappa_1/2$ inserted at $0, 1, \infty$. We will also use other notations $1 - \kappa_1  = \alpha,\,1 -\kappa_2 = \beta$ to conform with the reference. The formula (\ref{z4}) is supposed to work when $2\kappa_1>\kappa_2$, which is Troyanov condition in the geometric language. Energy-momentum tensor is decomposed as
\begin{equation}
t^{(0)}(y) = \frac{\delta_2}{y^2}  + \frac{\delta_1}{(y-1)^2} - \frac{\delta_2}{y(y-1)},\, t^{(1)}(y)= \frac{\epsilon_2}{(y-z)^2} + \frac{x(x-1)c}{y(y-1)(y-z)} - \frac{\epsilon_2}{y(y-1)}
\end{equation}
For the basis chosen in \cite{Alkalaev_2019}, monodromy matrix around zero is $-\text{diag }(e^{i\pi \beta}, e^{-i\pi \beta})+0 \epsilon_2 + \dots$; around $z$ we have 
\begin{equation}
M_{\gamma_z} = 1_{2 \times 2}+ i\epsilon_2
\begin{pmatrix}
I_{++} & I_{+-} \\
I_{-+} & I_{--}\\
\end{pmatrix}
\end{equation}
\begin{equation}
I_{++} = -I_{--} =  A  F_+ F_- (1-z)^\alpha (C z(1-z) + B(z))
\end{equation}
\begin{equation}
I_{+-} = - A  (1-z)^\alpha z^{-\beta} F_-^2 (C z(1-z) +D_-(z));\, I_{-+} = A (1-z)^\alpha z^{\beta} F_+^2 (C z(1-z) +D_+(z));\,
\end{equation}
where
\begin{align}
&A = \frac{2\pi^2}{\sin \pi \beta},\,B(z) = 1-z(\alpha+2) +z(1-z) \frac{d \log(F_+ F_-)}{dz}; \nonumber \\
&D_{\pm} = 1-z(\alpha+2) + z(1-z) \frac{d \log F_\pm^2}{dz}  \pm \beta(1-z)
\end{align}
and
\begin{equation}
F_\pm(z) = {}_{2}F_{1}(\frac{1 \pm \beta}{2}, \frac{1 \pm \beta}{2} + \alpha, 1 \pm \beta, z)
\end{equation}
In 0th order both these matrices are from $SU(2)$; to satisfy this condition for monodromy around $1$, matrix $J_0$ again needs to have a diagonal form $\text{diag }(a,1/a)$, but now $a$ is fixed at zeroth order: 
\begin{equation}
a^2 = -\frac{16^{-\beta } \Gamma \left(-\frac{\beta }{2}\right)^2 \Gamma \left(\frac{1}{2} (-2 \alpha +\beta +1)\right) \Gamma \left(\frac{1}{2} (2 \alpha +\beta +1)\right)}{\Gamma \left(\frac{\beta }{2}\right)^2 \Gamma \left(-\alpha -\frac{\beta }{2}+\frac{1}{2}\right) \Gamma \left(\alpha -\frac{\beta }{2}+\frac{1}{2}\right)}
\end{equation}
We can check that $a$ is real in the region where we expect positive curvature metrics to exist, i.e. Gauss-Bonnet + Troyanov condition are satisfied: $2\kappa_1 +\kappa_2<2,\,2\kappa_1>\kappa_2$. Now we need to find $C$ such that (\ref{su2cond}) is satisfied for monodromy around $z$. This condition in matrix form reads
\begin{equation}
\begin{pmatrix}
a I_{++} & a I_{+-} \\
a^{-1} I_{-+} & a^{-1} I_{--} \\
\end{pmatrix} = 
\begin{pmatrix}
a \br{I}_{++} & a^{-1} \br{I}_{-+} \\
a \br{I}_{+-} & a^{-1} \br{I}_{--} \\
\end{pmatrix}
\end{equation}
which reduces to 2 equations on $C \equiv c/\epsilon_2$ and $\br{C}$
\begin{equation}
C z(1-z) + B = \frac{\br{F}_+ \br{F}_-}{F_+ F_-} \frac{(1-\br{z})^\alpha}{(1-z)^\alpha} (\br{C} \br{z}(1-\br{z}) + \br{B})
\end{equation}
\begin{equation}
C z(1-z) + D_- = - \frac{1}{a^2} \frac{(1-\br{z})^\alpha}{(1-z)^\alpha} (z \br{z})^\beta \frac{\br{F}_+^2}{F_-^2} \left(\br{C} \br{z}(1-\br{z}) + \br{D}_+ \right)
\end{equation}
From these equations it follows
\begin{equation}
\br{C} = -\frac{1}{\br{z}(1-\br{z})} \left(\frac{1-z}{1-\br{z}} \right)^\alpha \frac{1}{\frac{\br{F}_+ \br{F}_-}{F_+ F_-}  + \frac{ (z \br{z})^\beta}{a^2} \frac{\br{F}_+^2}{F_-^2}} \cdot \left(D_- + \frac{(z \br{z})^\beta}{a^2} \frac{(1-\br{z})^\alpha}{(1-z)^\alpha}  \frac{\br{F}_+^2}{F_-^2}  \br{D}_+ - B + \frac{\br{F}_+ \br{F}_-}{F_+ F_-} \frac{(1-\br{z})^\alpha}{(1-z)^\alpha}  \br{B} \right) \label{cbar}
\end{equation}
Consider the asymptotics of this functions at $z=0,1,\infty$. First, at $z=0$ we have $D_\pm = 1 \pm \beta$,\,$B = \br{B} = 1$ and 
\begin{equation}
\br{C} \approx -\frac{1}{\br{z}} \left(1-\beta -1 + 1\right),\,z \to 0
\end{equation}
At $z \to 1$ leading asymptotic for hypergeometric function reads
\begin{equation}
F_\pm \approx (1-z)^{-\alpha} \frac{\Gamma(\alpha) \Gamma (\pm\beta +1)}{\Gamma \left(\frac{\pm \beta +1}{2}\right) \Gamma \left(\alpha +\frac{1\pm \beta}{2}\right)};\,\log F_\pm = -\alpha \log(1-z)
\end{equation}
Then,
\begin{equation}
B(z) \approx 1 - \alpha - 2 + 2 \alpha \frac{1-z}{1-z} = \alpha-1 = D_\pm
\end{equation}
the last bracket in (\ref{cbar}) reduces to
\begin{equation}
(\alpha-1) \left(\frac{1-z}{1-\br{z}} \right)^{-\alpha} \left( \frac{\br{F}_+ \br{F}_-}{F_+ F_-}  + \frac{ (z \br{z})^\beta}{a^2} \frac{\br{F}_+^2}{F_-^2}\right)
\end{equation}
and the asymptotic of $\br{C}$ is
\begin{equation}
\br{C} \approx  -\frac{1}{\br{z}-1} (1-\alpha),\,z\to 1
\end{equation}
Behaviour at infinity is found similarly; we obtain the asymptotic
\begin{equation}
    \br{C} \approx -\frac{1}{\br{z}} \left(1+\alpha\right),\,z \to \infty
\end{equation}
The sum of three residues of $c$, giving the volume, now becomes
\begin{equation}
\epsilon_2 (2\alpha+\beta-1) = \epsilon_2 (2-2\kappa_1-\kappa_2)
\end{equation}
consistent with MLG (\ref{z4}).

By these calculations we basically confirmed that accessory parameter calculated by perturbation theory in monodromy method has the same asymptotics as derivative of the classical solution $\pd \phi(z)$; i.e. correspondence with ``light'' limit.
\subsection{A comment on CFT derivation of previous results}
As was noted before, in the region of parameter space considered in this section numerical method of section \ref{sec3} does not work: non-trivial saddle point (dependent on both $x$ and $\br{x}$) in the integral over Liouville momentum $p$ disappears (in fact, in any channel). It is then reasonable to ask how do we expect to find nontrivially dependent on both $x, \br{x}$ Kähler potential from the CFT approach.

Apparently, to do this, one must account for ``discrete terms'', i.e. the residues at the poles of the structure constant that cross the integration contour $p \in \mathbb{R}$ in (\ref{decomp}) when we enter the domain $\eta_i + \eta_j<1/2$. The residue at each such pole has the form $\exp (f(x) + f(\br{x}))$; however, in the $b \to 0$ limit there is an infinite amount of these contributions (since the distance between the consecutive poles is $b$), among which there is no dominant one and all of them must be summed. After the summation, the correlator is no longer a product of holomorphic and antiholomorphic function and its logarithm can be a non-trivial Kähler potential.

In fact, the result of \cite{Balasubramanian_2017} mentioned in the beginning of section \ref{sec43} was reproduced from CFT in this reference from precisely these considerations. In that case it was possible, because classical conformal blocks with 2 heavy and 2 perturbatively heavy operators can be calculated exactly \cite{Fitzpatrick_2015}. It would be interesting to see if the same can be done in more general case, without ``perturbatively heavy'' approximation.
\section{Discussion} \label{sec5}
We finish with some interesting possible directions for the future work:
\begin{itemize}
    \item MLG correlators like (\ref{4pfmm}) have a simple enough form even without taking $p \to \infty$ limit to try and find some meaning of them. For the usual Weil-Petersson volumes finite-$p$ deformations were proposed and calculated in \cite{mertens2021} --- these are MLG boundary amplitudes. For ``sharp'' defects it can be shown that tachyon correlators that we examine coincide with these $p$-deformed volumes analytically continued to imaginary lengths (in the previous work \cite{Artemev_2022} it was mistakenly stated otherwise). It would be interesting to understand if it is possible to find an alternative definition of both of these objects, for example, in terms of representation theory of $U_q(sl_2)$, known to be relevant for Liouville gravity \cite{Fan_2022}, or from the point of view of quantization of Teichmuller space \cite{teschner2003liouville}. 
    
    \item One can consider the following alternative way to check the results of these article. By definition, MLG correlator we started with is an integral over $x$ of the product of Liouville and minimal model four-point functions. If after integration we obtain the moduli space volume in the semiclassical limit, it is reasonable to assume that the product of correlators reduces to a volume form of ZT metric. The Liouville 4-point function is expressed using formula (\ref{decomp}); in the semiclassical limit it might be possible to replace minimal model with $c<1$ Liouville theory (passing to the so-called ``generalized minimal gravity'') and then the same is valid for the matter correlator. Evaluating both integrals over Liouville momenta with saddle point method, we would find that leading exponential factors $\exp (\mp \frac{1}{b^2} S^{(4)} (p_\text{saddle},x,x) $ (``classical action'') cancel between Liouville and matter and only $O(1/c)$ corrections to classical conformal blocks and structure constants are left. Also, the determinant $(\pd^2 S^{(4)}/\pd p^2)^{-1} \mid_{p = p_{\text{saddle}}}$ appears after Gaussian integration in leading order; combining these 3 factors, one should obtain an alternative representation of the ZT metric volume form. If one could systematically compute $O(1/c)$ corrections to conformal block, this alternative proof may be possible to carry out numerically. 
    \item Generally the idea to use methods and exact results in classical and quantum Liouville CFT to study classical geometry of moduli spaces and JT gravity (which is intimately connected with it) seems very promising. One particular direction of thought is the geometric meaning of ``heavy'' degenerate operators in classical Liouville theory ($V_{1,n}$ with $n \sim 1/b^2$) and a related question of JT limit for ground ring operators (see e.g. \cite{wit1992}, \cite{Belavin:2005jy}) in MLG. We hope to obtain some insight into this question in the future.
    
    \item While this work was being prepared, another proposal \cite{eberhardt20232d} for the measure on moduli spaces of surfaces with conical defects appeared, which also agrees with semiclassical limit of MLG answers. It would be interesting to connect the two approaches. 
    We note one particular connection: an approximate expression (\ref{zn}) that we motivated from classical Liouville theory can be considered as following from  ``string'' and ``dilaton'' equations proven in \cite{eberhardt20232d} (equations $(1.3\text{a})$ and $(1.3\text{b})$). Indeed, for surfaces without geodesic boundaries from these equations it follows that moduli space volume for surfaces with one defect of very small deficit angle $\kappa \to 0$ vanishes linearly in $\kappa$ with the coefficient proportional to the power of Euler characteristic. The  arguments for ``light limit'' at section \ref{lightlim} can also be straightforwardly developed to obtain precisely the ``dilaton'' equation in a more general setting.
\end{itemize}
\section{Acknowledgements}
The author is grateful to Alexey Litvinov, Andrei Grigorev and Igor Chaban for stimulating discussions.  This work was supported by the Russian Science Foundation grant (project no. 23-12-00333).

\bibliographystyle{JHEP}
\bibliography{mlg2}

\providecommand{\href}[2]{#2}\begingroup\raggedright\begin{thebibliography}{10}

\bibitem{Knizhnik:1988ak}
V.~G. Knizhnik, A.~M. Polyakov and A.~B. Zamolodchikov, \emph{{Fractal
  Structure of 2D Quantum Gravity}},
  \href{http://dx.doi.org/10.1142/S0217732388000982}{\emph{Mod. Phys. Lett. A}
  {\bf 3} (1988) 819}.

\bibitem{DISTLER1989509}
J.~Distler and H.~Kawai, \emph{{Conformal field theory and 2D quantum
  gravity}},
  \href{http://dx.doi.org/https://doi.org/10.1016/0550-3213(89)90354-4}{\emph{Nuclear
  Physics B} {\bf 321} (1989) 509--527}.

\bibitem{mertens2022solvable}
T.~G. Mertens and G.~J. Turiaci, \emph{{Solvable Models of Quantum Black Holes:
  A Review on Jackiw-Teitelboim Gravity}},
  \href{http://arxiv.org/abs/2210.10846}{{\tt 2210.10846}}.

\bibitem{saad2019jt}
P.~Saad, S.~H. Shenker and D.~Stanford, \emph{{JT gravity as a matrix
  integral}},  \href{http://arxiv.org/abs/1903.11115}{{\tt 1903.11115}}.

\bibitem{mertens2021}
T.~G. Mertens and G.~J. Turiaci, \emph{{Liouville quantum gravity --
  holography, JT and matrices}},
  \href{http://dx.doi.org/10.1007/JHEP01(2021)073}{\emph{JHEP} {\bf 01} (2021)
  073}, [\href{http://arxiv.org/abs/2006.07072}{{\tt 2006.07072}}].

\bibitem{Zamolodchikov:2005fy}
A.~B. Zamolodchikov, \emph{{Three-point function in the minimal Liouville
  gravity}}, \href{http://dx.doi.org/10.1007/s11232-005-0003-3}{\emph{Theor.
  Math. Phys.} {\bf 142} (2005) 183--196},
  [\href{http://arxiv.org/abs/hep-th/0505063}{{\tt hep-th/0505063}}].

\bibitem{Belavin:2005jy}
A.~A. Belavin and A.~B. Zamolodchikov, \emph{{Integrals over moduli spaces,
  ground ring, and four-point function in minimal Liouville gravity}},
  \href{http://dx.doi.org/10.1007/s11232-006-0075-8}{\emph{Theor. Math. Phys.}
  {\bf 147} (2006) 729--754}.

\bibitem{turiaci2021}
G.~J. Turiaci, M.~Usatyuk and W.~W. Weng, \emph{{2D dilaton-gravity,
  deformations of the minimal string, and matrix models}},
  \href{http://dx.doi.org/10.1088/1361-6382/ac25df}{\emph{Class. Quant. Grav.}
  {\bf 38} (2021) 204001}, [\href{http://arxiv.org/abs/2011.06038}{{\tt
  2011.06038}}].

\bibitem{Artemev_2022}
A.~Artemev, \emph{Note on large-p limit of (2,2p+1) minimal liouville gravity
  and moduli space volumes},
  \href{http://dx.doi.org/10.1016/j.nuclphysb.2022.115876}{\emph{Nuclear
  Physics B} {\bf 981} (aug, 2022) 115876}.

\bibitem{Zograf_1988}
P.~G. Zograf and L.~A. Takhtadzhyan, \emph{{On Liouville's equation, accessory
  parameters, and the geometry of Teichmuller space for Riemann surfaces of
  genus 0}},
  \href{http://dx.doi.org/10.1070/SM1988v060n01ABEH003160}{\emph{Mathematics of
  the USSR-Sbornik} {\bf 60} (feb, 1988) 143}.

\bibitem{takhtajan2001hyperbolic}
L.~Takhtajan and P.~Zograf, \emph{{Hyperbolic 2-spheres with conical
  singularities, accessory parameters and Kähler metrics on
  $\mathcal{M}_{0,n}$}},  \href{http://arxiv.org/abs/math/0112170}{{\tt
  math/0112170}}.

\bibitem{bpz}
A.~A. Belavin, A.~M. Polyakov and A.~B. Zamolodchikov, \emph{{Infinite
  Conformal Symmetry in Two-Dimensional Quantum Field Theory}},
  \href{http://dx.doi.org/10.1016/0550-3213(84)90052-X}{\emph{Nucl. Phys. B}
  {\bf 241} (1984) 333--380}.

\bibitem{dornotto1994}
H.~Dorn and H.-J. Otto, \emph{{Two- and three-point functions in Liouville
  theory}}, \href{http://dx.doi.org/10.1016/0550-3213(94)00352-1}{\emph{Nuclear
  Physics B} {\bf 429} (Oct, 1994) 375–388}.

\bibitem{zamzam1996}
A.~Zamolodchikov and A.~Zamolodchikov, \emph{{Conformal bootstrap in Liouville
  field theory}},
  \href{http://dx.doi.org/10.1016/0550-3213(96)00351-3}{\emph{Nuclear Physics
  B} {\bf 477} (Oct, 1996) 577–605}.

\bibitem{seibergnotes}
N.~Seiberg, \emph{{Notes on Quantum Liouville Theory and Quantum Gravity}},
  \href{http://dx.doi.org/10.1143/PTP.102.319}{\emph{Progress of Theoretical
  Physics Supplement} {\bf 102} (03, 1990) 319--349},
  [\href{http://arxiv.org/abs/https://academic.oup.com/ptps/article-pdf/doi/10.1143/PTP.102.319/5376238/102-319.pdf}{{\tt
  https://academic.oup.com/ptps/article-pdf/doi/10.1143/PTP.102.319/5376238/102-319.pdf}}].

\bibitem{Hadasz_2004}
L.~Hadasz and Z.~Jask{\'{o}}lski, \emph{{Classical Liouville action on the
  sphere with three hyperbolic singularities}},
  \href{http://dx.doi.org/10.1016/j.nuclphysb.2004.03.012}{\emph{Nuclear
  Physics B} {\bf 694} (aug, 2004) 493--508}.

\bibitem{Hadasz_2003}
L.~Hadasz and Z.~Jask{\'{o}}lski, \emph{Polyakov conjecture for hyperbolic
  singularities},
  \href{http://dx.doi.org/10.1016/j.physletb.2003.08.075}{\emph{Physics Letters
  B} {\bf 574} (nov, 2003) 129--135}.

\bibitem{Cantini_2001}
L.~Cantini, P.~Menotti and D.~Seminara, \emph{{Proof of Polyakov conjecture for
  general elliptic singularities}},
  \href{http://dx.doi.org/10.1016/s0370-2693(01)00998-4}{\emph{Physics Letters
  B} {\bf 517} (sep, 2001) 203--209}.

\bibitem{franc1995}
P.~Francesco, P.~Ginsparg and J.~Zinn-Justin, \emph{2d gravity and random
  matrices},
  \href{http://dx.doi.org/10.1016/0370-1573(94)00084-g}{\emph{Physics Reports}
  {\bf 254} (Mar, 1995) 1–133}.

\bibitem{Moore:1991ir}
G.~W. Moore, N.~Seiberg and M.~Staudacher, \emph{{From loops to states in 2-D
  quantum gravity}},
  \href{http://dx.doi.org/10.1016/0550-3213(91)90548-C}{\emph{Nucl. Phys. B}
  {\bf 362} (1991) 665--709}.

\bibitem{belzam2009}
A.~A. Belavin and A.~B. Zamolodchikov, \emph{{On Correlation Numbers in 2D
  Minimal Gravity and Matrix Models}},
  \href{http://dx.doi.org/10.1088/1751-8113/42/30/304004}{\emph{J. Phys. A}
  {\bf 42} (2009) 304004}, [\href{http://arxiv.org/abs/0811.0450}{{\tt
  0811.0450}}].

\bibitem{alesh2016}
K.~Aleshkin and V.~Belavin, \emph{{On the construction of the correlation
  numbers in Minimal Liouville Gravity}},
  \href{http://dx.doi.org/10.1007/JHEP11(2016)142}{\emph{JHEP} {\bf 11} (2016)
  142}, [\href{http://arxiv.org/abs/1610.01558}{{\tt 1610.01558}}].

\bibitem{mazzeo2015teichmuller}
R.~Mazzeo and H.~Weiss, \emph{Teichm\"uller theory for conic surfaces},
  \href{http://arxiv.org/abs/1509.07608}{{\tt 1509.07608}}.

\bibitem{tarn2011}
G.~Tarnopolsky, \emph{{Five-point Correlation Numbers in One-Matrix Model}},
  \href{http://dx.doi.org/10.1088/1751-8113/44/32/325401}{\emph{J. Phys. A}
  {\bf 44} (2011) 325401}, [\href{http://arxiv.org/abs/0912.4971}{{\tt
  0912.4971}}].

\bibitem{Fateev_2008}
V.~A. Fateev and A.~V. Litvinov, \emph{{Multipoint correlation functions in
  Liouville field theory and minimal Liouville gravity}},
  \href{http://dx.doi.org/10.1007/s11232-008-0038-3}{\emph{Theoretical and
  Mathematical Physics} {\bf 154} (mar, 2008) 454--472}.

\bibitem{Hadasz_2005}
L.~Hadasz, Z.~Jask{\'{o}}lski and M.~Piątek, \emph{{Classical geometry from
  the quantum Liouville theory}},
  \href{http://dx.doi.org/10.1016/j.nuclphysb.2005.07.003}{\emph{Nuclear
  Physics B} {\bf 724} (sep, 2005) 529--554}.

\bibitem{harrison2022liouville}
S.~M. Harrison, A.~Maloney and T.~Numasawa, \emph{{Liouville Theory and the
  Weil-Petersson Geometry of Moduli Space}},
  \href{http://arxiv.org/abs/2210.08098}{{\tt 2210.08098}}.

\bibitem{fırat2023hyperbolic}
A.~H. Fırat, \emph{Hyperbolic string tadpole},
  \href{http://arxiv.org/abs/2306.08599}{{\tt 2306.08599}}.

\bibitem{teschner2014supersymmetric}
J.~Teschner, \emph{{Supersymmetric gauge theories, quantisation of moduli
  spaces of flat connections, and Liouville theory}},
  \href{http://arxiv.org/abs/1412.7140}{{\tt 1412.7140}}.

\bibitem{Zamolodchikov1987ConformalSI}
A.~B. Zamolodchikov, \emph{Conformal symmetry in two-dimensional space:
  Recursion representation of conformal block}, {\emph{Theoretical and
  Mathematical Physics} {\bf 73} (1987) 1088--1093}.

\bibitem{Harlow_2011}
D.~Harlow, J.~Maltz and E.~Witten, \emph{{Analytic continuation of Liouville
  theory}}, \href{http://dx.doi.org/10.1007/jhep12(2011)071}{\emph{Journal of
  High Energy Physics} {\bf 2011} (dec, 2011) }.

\bibitem{Alkalaev_2020}
K.~Alkalaev and M.~Pavlov, \emph{Holographic variables for {CFT}2 conformal
  blocks with heavy operators},
  \href{http://dx.doi.org/10.1016/j.nuclphysb.2020.115018}{\emph{Nuclear
  Physics B} {\bf 956} (jul, 2020) 115018}.

\bibitem{trufpopova}
private communication~with J.~Popova and A.~Trufanov.

\bibitem{Balasubramanian_2017}
V.~Balasubramanian, A.~Bernamonti, B.~Craps, T.~D. Jonckheere and F.~Galli,
  \emph{{Heavy-heavy-light-light correlators in Liouville theory}},
  \href{http://dx.doi.org/10.1007/jhep08(2017)045}{\emph{Journal of High Energy
  Physics} {\bf 2017} (aug, 2017) }.

\bibitem{Alkalaev_2019}
K.~Alkalaev and M.~Pavlov, \emph{Four-point conformal blocks with three heavy
  background operators},
  \href{http://dx.doi.org/10.1007/jhep08(2019)038}{\emph{Journal of High Energy
  Physics} {\bf 2019} (aug, 2019) }.

\bibitem{Fitzpatrick_2015}
A.~L. Fitzpatrick, J.~Kaplan and M.~T. Walters, \emph{Virasoro conformal blocks
  and thermality from classical background fields},
  \href{http://dx.doi.org/10.1007/jhep11(2015)200}{\emph{Journal of High Energy
  Physics} {\bf 2015} (nov, 2015) }.

\bibitem{Fan_2022}
Y.~Fan and T.~G. Mertens, \emph{{From quantum groups to Liouville and dilaton
  quantum gravity}},
  \href{http://dx.doi.org/10.1007/jhep05(2022)092}{\emph{Journal of High Energy
  Physics} {\bf 2022} (may, 2022) }.

\bibitem{teschner2003liouville}
J.~Teschner, \emph{{From Liouville Theory to the Quantum Geometry of Riemann
  Surfaces}},  \href{http://arxiv.org/abs/hep-th/0308031}{{\tt
  hep-th/0308031}}.

\bibitem{wit1992}
E.~Witten, \emph{Ground ring of two-dimensional string theory},
  \href{http://dx.doi.org/10.1016/0550-3213(92)90454-j}{\emph{Nuclear Physics
  B} {\bf 373} (Mar, 1992) 187–213}.

\bibitem{eberhardt20232d}
L.~Eberhardt and G.~J. Turiaci, \emph{{2D dilaton gravity and the
  Weil-Petersson volumes with conical defects}},
  \href{http://arxiv.org/abs/2304.14948}{{\tt 2304.14948}}.

\end{thebibliography}\endgroup
\end{document}